\documentclass[aps,twocolumn,nofootinbib,superscriptaddress,PRD]{revtex4}

\usepackage{amsmath,bm}
\usepackage{graphicx}
\usepackage{epsfig}
\usepackage{multirow}
\usepackage{color}
\usepackage[breaklinks,colorlinks,urlcolor=blue,citecolor=blue,linkcolor=blue]{hyperref}
\usepackage{verbatim}
\begin{document}

\title{Particle production at large $p_{\rm T}$ in Xe+Xe collisions \\with jet quenching using the higher twist approach}

\date{\today  \hspace{1ex}}

\author{Qing Zhang}
\affiliation{Key Laboratory of Quark \& Lepton Physics (MOE) and Institute of Particle Physics, Central China Normal University, Wuhan 430079, China}

\author{Wei Dai}
\affiliation{School of Mathematics and Physics, China University of Geosciences (Wuhan), Wuhan 430074, China}

\author{Lei Wang}
\affiliation{Key Laboratory of Quark \& Lepton Physics (MOE) and Institute of Particle Physics, Central China Normal University, Wuhan 430079, China}

\author{Ben-Wei Zhang}
\email{bwzhang@mail.ccnu.edu.cn}
\affiliation{Key Laboratory of Quark \& Lepton Physics (MOE) and Institute of Particle Physics, Central China Normal University, Wuhan 430079, China}
\affiliation{Guangdong Provincial Key Laboratory of Nuclear Science, Institute of Quantum Matter, South China Normal University, Guangzhou 510006, China}

\author{Enke Wang}
\affiliation{Guangdong Provincial Key Laboratory of Nuclear Science, Institute of Quantum Matter, South China Normal University, Guangzhou 510006, China}
\affiliation{Guangdong-Hong Kong Joint Laboratory of Quantum Matter, Southern Nuclear Science Computing Center, South China Normal University, Guangzhou 510006, China}
\affiliation{Key Laboratory of Quark \& Lepton Physics (MOE) and Institute of Particle Physics, Central China Normal University, Wuhan 430079, China}

\begin{abstract}
The production of $\pi^0$, $\eta$, and $\phi$ in the most central (0\%-5\%) Xe+Xe collisions at $\sqrt{s_{NN}}$ = 5.44~TeV is investigated in the framework of the perturbative QCD (pQCD) improved parton model at an accuracy of next-to-leading order (NLO). The jet quenching effect is effectively incorporated by medium-modified fragmentation functions via the higher-twist approach. Predictions of the nuclear modification factors of $\pi^0$, $\eta$, and $\phi$ as functions of the transverse momentum $p_{\rm T}$ are made with the jet transport parameter $\hat{q}_0$, which is extracted from the available experimental data of charged hadrons provided by ALICE and CMS. The particle ratios $\eta/\pi^0,\ \phi/\pi^0$ as functions of $p_{\rm T}$ in Xe+Xe collisions at $\sqrt{s_{NN}}$ = 5.44~TeV as well as in 0\%-5\% Pb+Pb collisions at $\sqrt{s_{NN}}$ = 5.02~TeV are also presented. The numerical simulations of the scaled ratios of charged hadron production in the Xe+Xe 5.44~TeV system over those in the Pb+Pb 5.02~TeV system give a good description of the CMS data, and the scaled ratios of $\pi^0$, $\eta$, and $\phi$ production coincide with the curve of charged hadron production.
% $\sqrt{s_\mathrm{NN}}$


{\bf DOI: }10.1088/1674-1137/ac7b75
\end{abstract}

\maketitle
%\begin{spacing}{2.3}
\section{Introduction}%\romannumeral1.\ 
\label{sec:intro}

In ultra-relativistic heavy-ion collisions (HIC) at the Relativistic Heavy-Ion Collider (RHIC) at Brookhaven National Laboratory (BNL) and the Large Hadron Collider (LHC) at the European Organization for Nuclear Research (CERN), strongly coupled QCD matter known as quark-gluon plasma (QGP) is created, which exhibits many intriguing properties. QCD partons produced from early stage collisions may traverse through QGP, and its interactions with other partons in the QGP may lead to the attenuation of its energy, that is, jet quenching~\cite{Wang:1991xy,Zhang:2003wk,Gyulassy:2003mc,Qin:2015srf,Ma:2010dv,Fochler:2011en}. Among the convincing evidence of jet quenching effects is the strong suppression of inclusive hadron spectra at high transverse momentum ($p_{\rm T}$)~\cite{Gyulassy:2003mc}. Abundant experimental data from the RHIC and LHC on identified-hadron yields help us better understand the processes of jet-medium interactions and are well described within the next-to-leading order (NLO) perturbative QCD (pQCD) improved parton model incorporated with the higher-twist approach~\cite{Zhang:2003wk,Guo:2000nz,Chen:2010te,Chen:2011vt,Zhang:2003yn,Schafer:2007xh,Liu:2015vna,Kidonakis:2000gi}. Studies on the medium modification effect on different final-state hadron production in various collision systems are therefore essential for constraining our understanding of hadron suppression patterns~\cite{Dai:2015dxa,Dai:2017tuy,Dai:2017piq,Ma:2018swx}.

Suppression represented by the nuclear modification factor $R_{AA}$ of different final-state hadrons, $\pi^0$, $\eta$, and $\phi$, at large $p_{\rm T}$~\cite{PHENIX:2006ujp,PHENIX:2010hvs,STAR:2011iap} provides useful information in many respects, such as extracting the jet transport coefficient $\hat{q}$~\cite{JET:2013cls}. Moreover, the particle ratios $\eta/\pi^0$, $\rho^0/\pi^0$, and $\phi/\pi^0$ can help better understand energy-loss patterns. In our previous research on the production of different final-state hadrons in HIC~\cite{Dai:2015dxa,Dai:2017tuy,Dai:2017piq,Ma:2018swx}, we concluded that the leading hadron productions in HIC are the combined results of three factors: the initial hard parton-jet spectrum, the parton energy loss mechanism, and parton fragmentation functions (FFs) to the hadron in vacuum. For instance, the derived yield ratios $\eta/\pi^0$ and $\rho^0/\pi^0$ in $p$+$p$ and $A$+$A$ collisions coincide at large $p_{\rm T}$. This is due to the fact that $\eta,\ \pi^0,\ \rho^0$ are all dominated by quark fragmentation contributions at very large $p_{\rm T}$ in $p$+$p$ collisions, and the jet quenching effect will enhance the quark fragmentation contribution fraction (with a relatively weak $p_{\rm T}$ and $z_h$ dependence on their quark FFs, where $z_h$ denotes the momentum fraction of the hadron $h$ fragmented from a scattered quark or gluon). Therefore, at very large $p_{\rm T}$ in $A$+$A$ collisions, the relative contribution of quark and gluon fragmentations is small, and the particle ratios $\eta/\pi^0,\ \rho^0/\pi^0$ in $A$+$A$ collisions will mainly be determined by the ratios of quark FFs to the different final-state hadrons, which is the same as that in $p$+$p$ collisions. For the mesons $\phi$ and $\omega$, gluon fragmentation contributions dominate at large $p_{\rm T}$ in $p$+$p$ collisions, and the particle ratios $\phi/\pi^0$ and $\omega/\pi^0$ in $A$+$A$ collisions vary from those in $p$+$p$ collisions. The magnitude of this variation can help expose the difference between quark and gluon energy loss. Hence, in this study, we choose the $\pi^0,\ \eta$, and $\phi$ mesons as benchmarks to display final-state hadron yields in HIC. To achieve this, it is important to first investigate the amount of this variation in different nuclear-nuclear collision systems, which will introduce different energy densities and path lengths in the QGP medium. The emergence of experimental measurements in Xe+Xe collisions at $\sqrt{s_{NN}}$ = 5.44~TeV can facilitate such investigations~\cite{Xie:2019oxg} because they are conducted at similar colliding energies with Pb+Pb collisions at $\sqrt{s_{NN}}$ = 5.02~TeV but have an intermediate-size collision system between previous proton-proton ($p$+$p$), $p$+Pb, and Pb+Pb collisions~\cite{CMS:2018yyx,ALICE:2018hza}.

The rest of this paper is organized as follows. The theoretical framework for leading hadron production in $p$+$p$ collisions is presented in Sec.~{\uppercase\expandafter{\romannumeral2}}, and the $p$+$p$ spectra of charged hadron, $\pi^0$, $\eta$, and $\phi$ production are plotted. In Sec.~{\uppercase\expandafter{\romannumeral3}}, we deliberate on the nuclear modifications of leading hadron yields due to the jet quenching effect in $A$+$A$ collisions. We subsequently investigate the nuclear modification factor $R_{AA}$ of  charged hadrons and $\pi^0$, $\eta$, and $\phi$ mesons as well as their yield ratios in Xe+Xe collisions at $\sqrt{s_{NN}}$ = 5.44~TeV. 
 %the cases in Pb+Pb collisions at $\sqrt{s_{NN}}$ = 5.02~TeV are given in supplement. 
In Sec.~{\uppercase\expandafter{\romannumeral4}}, the scaled ratios of different final-state hadron production in the Xe+Xe 5.44~TeV system over those in the Pb+Pb 5.02~TeV system, denoted as $R^{\rm Xe}_{\rm Pb}$, is demonstrated. A brief summary is given in Sec.~{\uppercase\expandafter{\romannumeral5}}.

\section{Leading hadron yields in \lowercase{\bf $p+p$} collisions}%\romannumeral2.\ 
\label{sec:pp}

%%%%%%%%%%%%%%%%%%%%%%%%%%%%%%%%%%%%%%%%%%
%%***********    
%%Figure.ppbaseline ****
%\hspace{0.7in}
\begin{figure*}[!t]
\begin{center}

	\begin{minipage}{1\linewidth}
		\begin{center}
			\resizebox{0.75\textwidth}{!}{
				\includegraphics{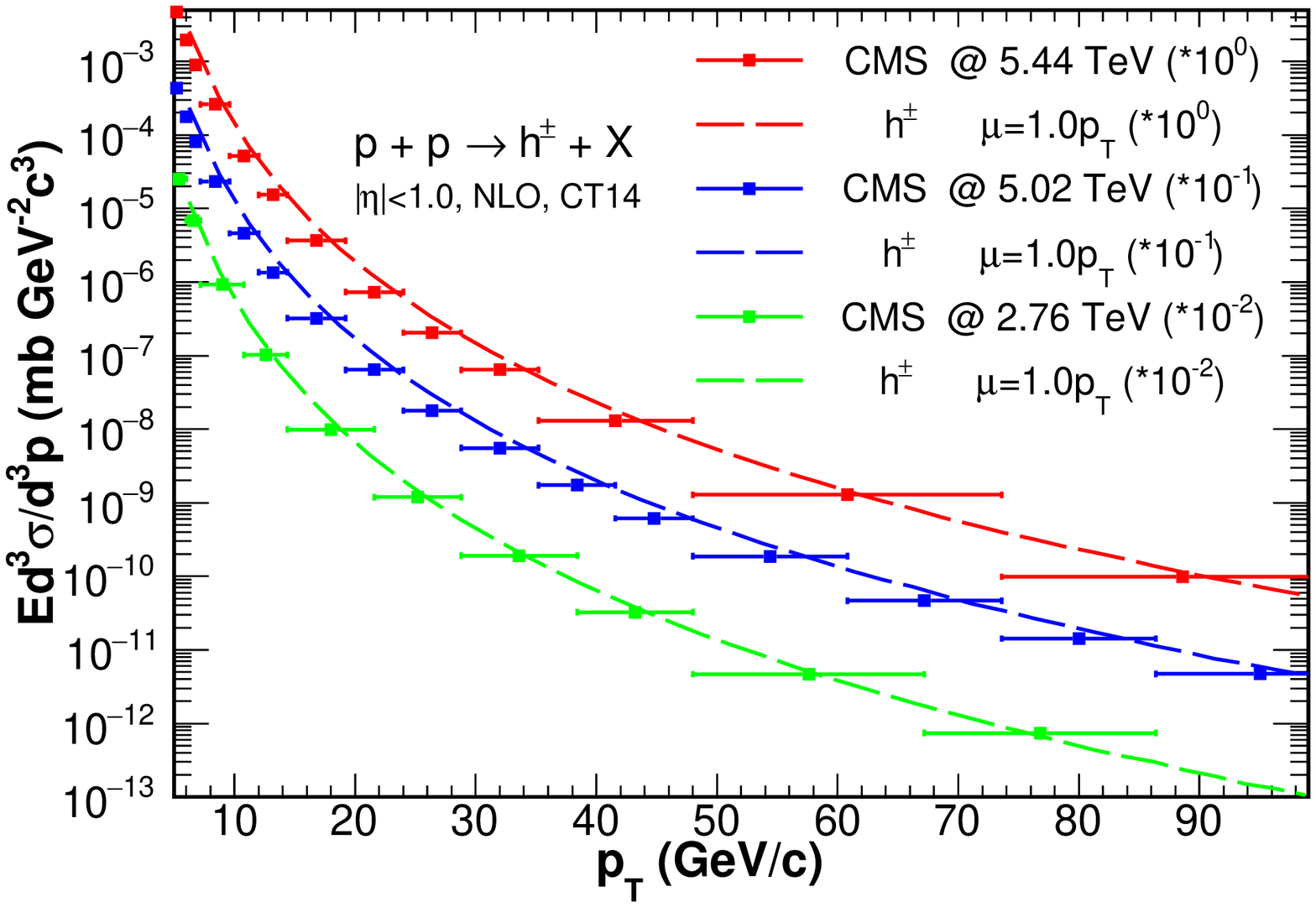}
				\includegraphics{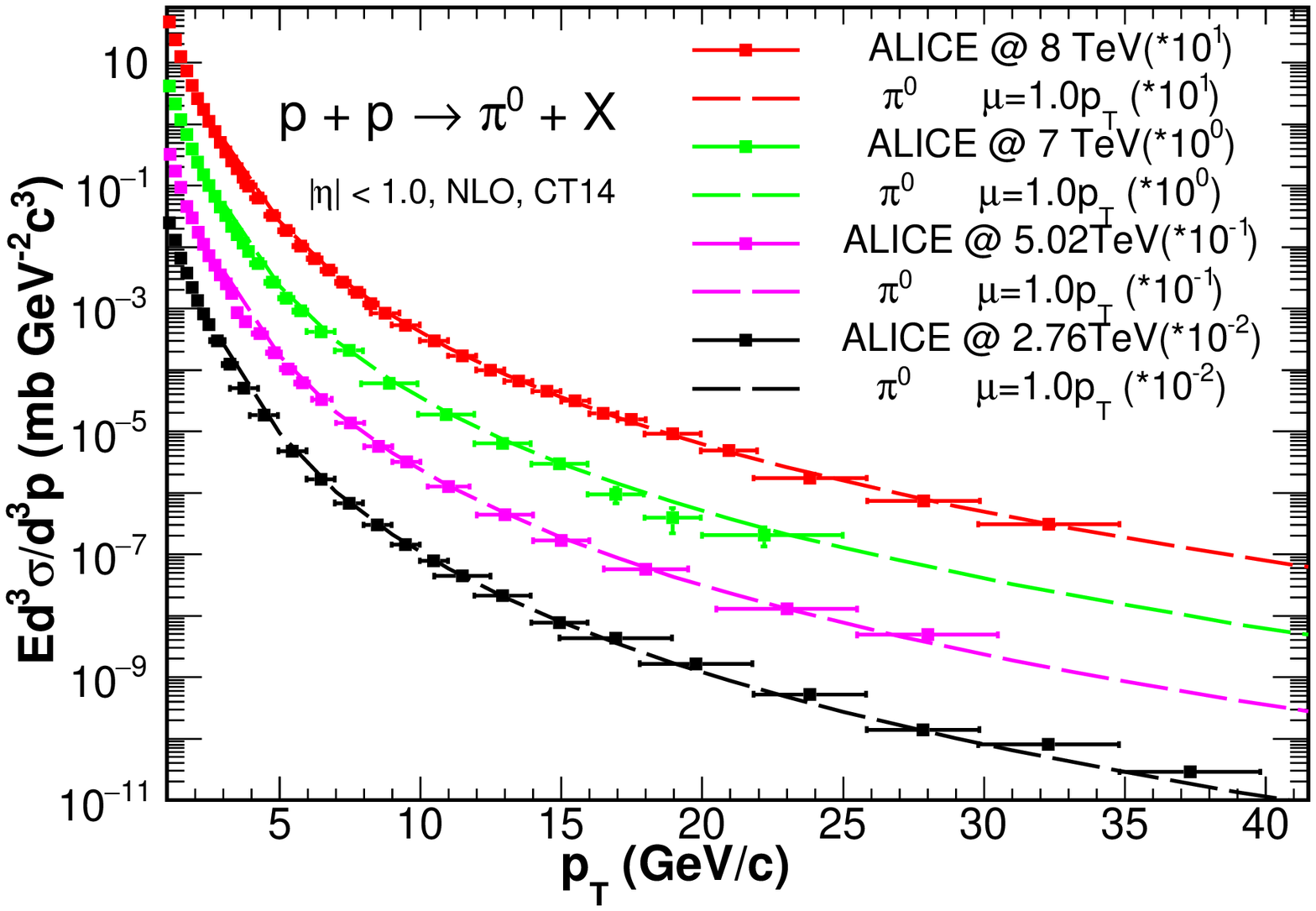}
			}
		\end{center}
	\end{minipage}
	
	\begin{minipage}{1\linewidth}
		\begin{center}
			\resizebox{0.75\textwidth}{!}{
				\includegraphics{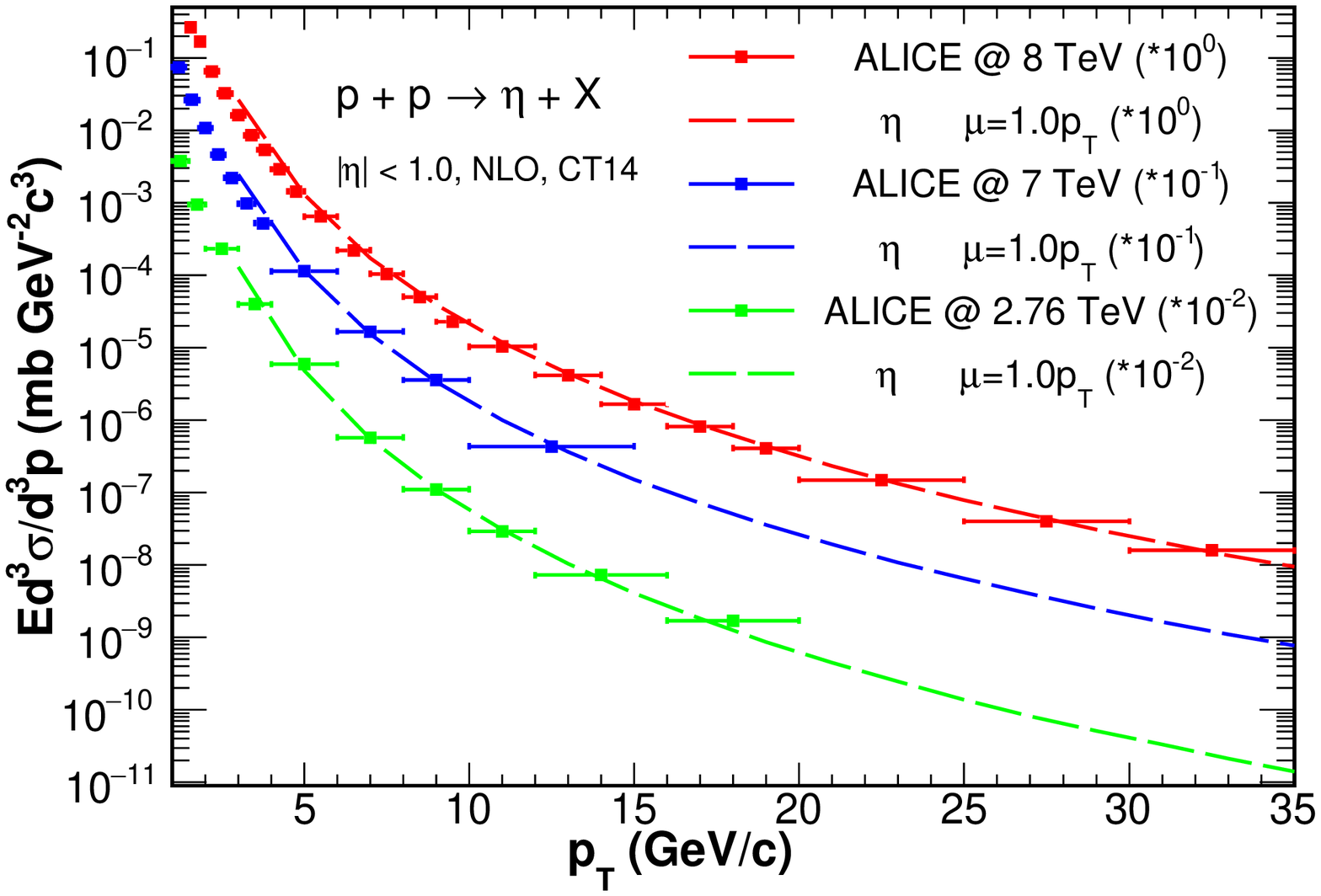}
				\includegraphics{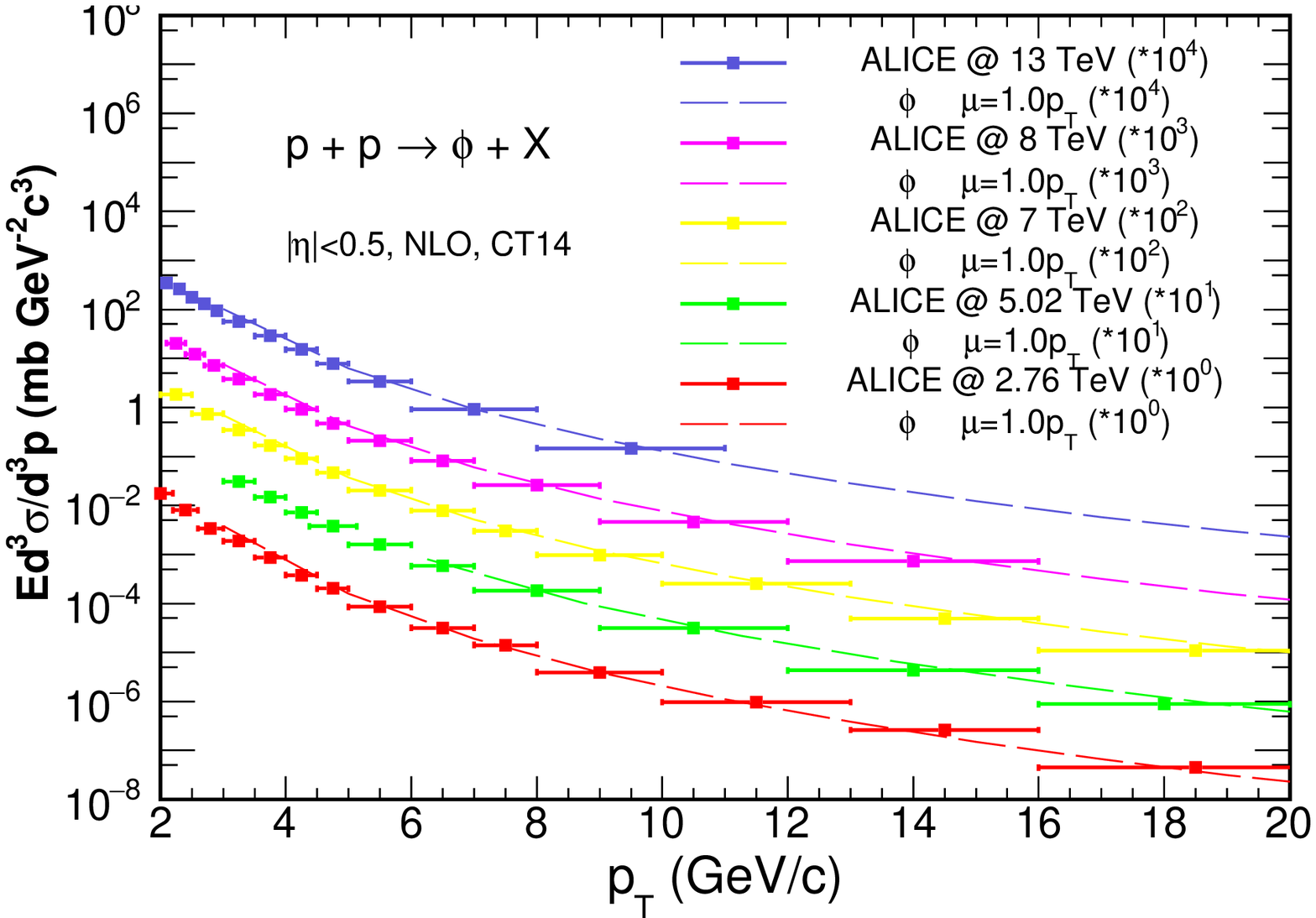}
			}
		\end{center}
	\end{minipage}

\hspace*{0.7in}
%\vspace*{-0.2in}
\caption{(color online) Final-state calculations of charged hadrons (upper left), $\pi^0$ (upper right), $\eta$ (bottom left), and $\phi$ (bottom right) in $p$+$p$ collisions over a wide range of LHC energy scales compared with available experimental data~\cite{Acharya:2017hyu,ALICE:2019hno,Abelev:2012cn,Acharya:2017tlv,CMS:2012aa,CMS:2016xef,CMS:2018yyx,Adam:2017zbf,ALICE:2021ptz,ALICE:2019hyb,ALICE:2020jsh}.}
\label{fig:ppbaseline}
\end{center}
\end{figure*}
%\hspace*{-0.5in}
%%%%%%%%%%%%%%%%%%%%%%%%%%%%%%%%%%%%%%%%%%%%%%%%%%%%%%%

Within the pQCD improved parton model at NLO~\cite{Kidonakis:2000gi}, the inclusive cross section of single hadron production in $p$+$p$ collisions is determined by two factors: the initial hard parton-jet spectrum $F_{q,g}(p_{\rm T}/z_h)$ and the parton FFs to final-state hadrons in vacuum, $D_{q,g\to h}(z_{h}, Q^2)$.
\begin{eqnarray}
	\frac{1}{p_{\rm T}}\frac{d\sigma_{h}}{dp_{\rm T}}=\int F_{q}(\frac{p_{\rm T}}{z_{h}})\cdot D_{q\to h}(z_{h}, Q^2)\frac{dz_{h}}{z_{h}^2} \nonumber  \\
	+ \int F_{g}(\frac{p_{\rm T}}{z_{h}})\cdot D_{g\to h}(z_{h}, Q^2)\frac{dz_{h}}{z_{h}^2}  \,\,\, .
	\label{eq:pp}
\end{eqnarray}
where $F_{q,g}(p_{\rm T}/z_h)$ is the convolution of initial parton distribution functions (PDFs) and partonic scattering cross sections, in which $z_h$ is the momentum fraction carried by the final hadron of its parent parton at the fragmentation scale $Q$. In this study, the factorization, renormalization, and fragmentation scales are taken to be equal and proportional to the final-state $p_{\rm T}$ of the leading hadron. CT14 parametrization~\cite{Dulat:2015mca} is employed for proton PDFs. KKP FFs~\cite{Kniehl:2000fe} are utilized for both $\pi^0$ and charged hadron production, AESSS FFs~\cite{Aidala:2010bn} are used for $\eta$ mesons, and NLO Dokshitzer-Gribov-Lipatov-Altarelli-Parisi (DGLAP) evolved FFs parametrized at the initial energy scale $Q_0^2=1.5$~GeV$^2$ by a broken SU(3) model are used for $\phi$ mesons~\cite{Indumathi:2011vn,Saveetha:2013jda,Hirai:2011si}.

Note that KKP parametrizations are chosen because during our calculations we find that the NLO theoretical results of $\pi^0$ production using the AKK08~\cite{Kniehl:2008et}, KKP~\cite{Kniehl:2000fe}, or KRE~\cite{Kretzer:2000yf} FFs cannot simultaneously describe the experimental data from relatively low collision energies, such as 200~GeV, at the RHIC to very high collisions energies, such as 13~TeV, at the LHC. This predicament is also pointed out in Refs.~\cite{deFlorian:2007aj,dEnterria:2013sgr,Metz:2016swz}, and the same occurs in the charged hadron and $\eta$ calculations. However, when we focus on LHC energies, KKP parametrizations can effectively describe the experimental data for charged hadron production. KKP parametrizations are also able to describe $\pi^0$ production at all collision energies within a margin of error with a re-scale factor $K$. The same method is applied to AESSS parametrizations in $\eta$ FFs to provide decent $p$+$p$ baselines at all collision energies with a $K$ factor.

In Fig.~\ref{fig:ppbaseline}, we present the numerical results of the final-state yields of charged hadrons, $\pi^0,\ \eta$, and $\phi$ at 2.76~TeV to 13~TeV and their comparison with all available experimental data~\cite{Acharya:2017hyu,ALICE:2019hno,Abelev:2012cn,Acharya:2017tlv,CMS:2012aa,CMS:2016xef,CMS:2018yyx,Adam:2017zbf,ALICE:2021ptz,ALICE:2019hyb,ALICE:2020jsh}. For charged hadron production, our calculation results agree well with experimental data~\cite{CMS:2012aa,CMS:2016xef,CMS:2018yyx} at all available energies in the margin of error when we fix the scales at $\mu = \mu_f = \mu_r = 1.0\ p_{\rm T}$. For the $\pi^0$ yields~\cite{Acharya:2017hyu,ALICE:2019hno,Abelev:2012cn,Acharya:2017tlv}, we utilize KKP parametrizations of FFs in vacuum with $K_{\pi^0}=0.5$ and scales of $\mu = 1.0\ p_{\rm T}$. For the $\eta$ yields~\cite{Acharya:2017hyu,ALICE:2019hno,Abelev:2012cn,Acharya:2017tlv}, AESSS FFs with the rescale factor $K_{\eta}=0.6$ are utilized when fixing $\mu = 1.0\ p_{\rm T}$. $\phi$ production with $\mu = 1.0\ p_{\rm T}$ also gives a decent description of the ALICE measurements~\cite{Adam:2017zbf,ALICE:2021ptz,ALICE:2019hyb,ALICE:2020jsh} as the collision energy reaches $13$~TeV.
%$\rho^0$:$\mu=0.5p_{\rm T}$

\section{$R_{AA}$ and particle ratios in X\lowercase{e}+X\lowercase{e} collisions}%\romannumeral3.\ 
\label{sec:AA}

To facilitate the parton energy loss mechanism of final-state hadron production in HIC, we factorize the process into two steps. A fast parton first loses energy owing to multiple scatterings with other partons in the hot and dense medium. It then fragments into final-state hadrons in vacuum. The total energy loss is carried away by radiated gluons and embodied in medium-modified quark FFs with the higher-twist approach~\cite{Zhang:2003wk,Guo:2000nz,Chen:2010te,Chen:2011vt,Zhang:2003yn,Schafer:2007xh},
\begin{eqnarray}
	\tilde{D}_{q}^{h}(z_h,Q^2) &=&
	D_{q}^{h}(z_h,Q^2)+\frac{\alpha_s(Q^2)}{2\pi}
	\int_0^{Q^2}\frac{d\ell_T^2}{\ell_T^2} \nonumber\\
	&&\hspace{-0.7in}\times \int_{z_h}^{1}\frac{dz}{z} \left[ \Delta\gamma_{q\rightarrow qg}(z,x,x_L,\ell_T^2)D_{q}^h(\frac{z_h}{z},Q^2)\right.
	\nonumber\\
	&&\hspace{-0.2 in}+ \left. \Delta\gamma_{q\rightarrow
		gq}(z,x,x_L,\ell_T^2)D_{g}^h(\frac{z_h}{z},Q^2) \right] .
	\label{eq:mmFFs}
\end{eqnarray}
where $\ell_T$ is the transverse momentum of the radiated gluons. Gluon radiation is then induced by the scattering of the quark with another gluon carrying a finite momentum fraction $x$. $x_L$ denotes the longitudinal momentum fraction, $z$ is the momentum fraction carried by the final quark, and $\Delta\gamma_{q\rightarrow qg}(z,x,x_L,\ell_T^2)$ and $\Delta\gamma_{q\rightarrow gq}(z,x,x_L,\ell_T^2) = \Delta\gamma_{q\rightarrow qg}(1-z,x,x_L,\ell_T^2)$ are the medium modified splitting functions~\cite{Guo:2000nz,Zhang:2003yn,Schafer:2007xh,Zhang:2003wk} which depend on the twist-four quark-gluon correlations inside the medium $T_{qg}^{A}(x,x_L)$,
\begin{eqnarray}
&&\Delta\gamma_{q\rightarrow qg}(z,x,x_L,\ell_T^2) \nonumber\\
&&\hspace{0.2in}=\Bigg[\frac{1+z^2}{(1-z)_{+}}T_{qg}^{A}(x,x_L) +\delta(1-z) \nonumber\\
&&\hspace{0.35in} \times\ \Delta T_{qg}^{A}(x,x_L)\Bigg] \frac{2\pi\alpha_sC_A}{\ell_T^2N_cf_q^A(x)};\\
\nonumber \\
&&\Delta\gamma_{q\rightarrow gq}(z,x,x_L,\ell_T^2)=\Delta\gamma_{q\rightarrow qg}(1-z,x,x_L,\ell_T^2). \,
 \label{eq:splt}
\end{eqnarray}
where $\alpha_s$ is the strong coupling constant, $f_q^A(x)$ is the initial hard parton-jet spectrum, and
\begin{eqnarray}
\frac{T^{A}_{qg}(x,x_L)}{f_q^A(x)} &=&\frac{N_{c}^{2}-1}{4\pi\alpha_sC_{R}}\frac{1+z}{2} \int dy^{-}
2 \sin^{2}\left[\frac{y^{-}\ell_{T}^{2}}{4Ez(1-z)}\right] \nonumber \\
&&\hspace{0.0 in}\times\left[\hat{q}_R(E,x_L,y)+c(x,x_{L}) \hat{q}_R(E,0,y)\right]. \,
 \label{eq:corr2}
\end{eqnarray}
is proportional to jet transport parameter $\hat{q}_R(E,y)$ when assuming $x \gg x_L,\,x_T$~\cite{Zhang:2003wk,Guo:2000nz,Chen:2010te,Chen:2011vt,Zhang:2003yn,Schafer:2007xh,Liu:2015vna}. The medium-modified FFs are averaged over the initial production position and jet propagation direction as follows~\cite{Chen:2010te,Chen:2011vt}:
\begin{eqnarray}
\langle \tilde D_{a}^{h}(z_h,Q^2,E,b)
\rangle &=& \frac{1}{\int
d^2{r}t_{A}(|\vec{r}|)t_{B}(|\vec{b}-\vec{r}|)} \nonumber \\
&&\hspace{-1.0in}\times \int\frac{d\phi}{2\pi}d^2{r}
t_{A}(|\vec{r}|)t_{B}(|\vec{b}-\vec{r}|)\tilde{D}_{a}^{h}(z_h,Q^2,E,r,\phi,b). \nonumber \\
\label{eq:frag}
\end{eqnarray}
where the impact parameter $b$ and nuclear thickness function $t_{A,B}$ are provided by the Glauber Model~\cite{Moller:2015fba,Zakharov:2018ctz}. Then, we assume the total energy loss is the energy carried away by the radiated gluon.
\begin{eqnarray}
\frac{\Delta E}{E} &=& \frac{2N_{c}\alpha_s}{\pi} \int dy^-dz
{d\ell_T^2}
\frac{1+z^2}{\ell_T^4} \nonumber \\
&& \times \left(1-\frac{1-z}{2}\right)\hat q(E,y)
\sin^2\left[\frac{y^-\ell_T^2}{4Ez(1-z)}\right], 
\label{eq:de}
\end{eqnarray}
which is also proportional to the jet transport parameter $\hat{q}_R(E,y)$. The jet transport parameter $\hat{q}_R(E,y)$ depends on the space-time evolution of the QCD medium, and in our study, it is described by a (3+1)D viscous hydrodynamic model CLVisc~\cite{Pang:2012he,Pang:2014ipa,Pang:2018zzo}. To take the initial-state cold nuclear matter effects into consideration, EPPS16 NLO nuclear PDFs (nPDFs)~\cite{Eskola:2016oht} are employed. Therefore, leading hadron production in $A$+$A$ collisions at NLO can be obtained in a similar way as in $p$+$p$ collisions so that NLO partonic cross sections are convoluted with NLO nuclear PDFs and are then convoluted with the effective medium-modified fragmentaion functions $\tilde{D}_{q}^{h}(z_h,Q^2)$.

The nuclear modification factor $R_{AA}$ for single hadron production is defined as the ratio of cross sections in $A$+$A$ collisions over that in $p$+$p$ collisions scaled by the averaged number of binary $N$+$N$ collisions at a certain impact parameter $b$,
\begin{eqnarray}
	R^b_{AB}(p_{\rm T}, y)=\frac{d\sigma_{AB}^h/dydp_{\rm T}}{\langle N_{\rm bin}^{AB}(b) \rangle d\sigma_{pp}^h/dydp_{\rm T}} \ .
	\label{eq:raa}
\end{eqnarray}
where $\langle N_{\rm bin}^{AB}(b) \rangle$ is calculated using the optical Glauber method with a deformed Fermi distribution~\cite{Moller:2015fba,Zakharov:2018ctz},
\begin{eqnarray}
    &&R=1.1A^{1/3}-0.656A^{-1/3}\ \rm{fm}\\
	&&\rho_A(r,\theta)=\frac{\rho_0}{1+exp(r-R_A(\theta))/a}\\
	\nonumber\\
	&&R_A(\theta)=R[1+\beta_2Y_{20}(\theta)+\beta_4Y_{40}(\theta)]
\end{eqnarray}
in which for $^{129}$Xe, $A$ = 129, $a$ = 0.68 fm, $\beta_2$ = 0.162, and $\beta_4$= -0.003, and $Y_{20}$ and $Y_{40}$ are spherical harmonics~\cite{Moller:2015fba}.
The theoretical results of $R_{AA}$ for $\eta,\ \rho^0,\ \phi,\ \omega,\ K^0_{\rm S}$ production in central Pb+Pb collisions at $\sqrt{s_{NN}}$ = 2.76~TeV are in good agreement with experimental data~\cite{Dai:2015dxa,Dai:2017tuy,Dai:2017piq,Ma:2018swx}.

%%%%%%%%%%%%%%%%%%%%%%%%%%%%%%%%%%%%%%%%%%
%%***********    
%%Figure.chi2 ****
\begin{figure}[htbp]
	
 	\begin{minipage}{1\linewidth}
	    \begin{center}
		    \resizebox{0.8\textwidth}{!}{
		        \includegraphics{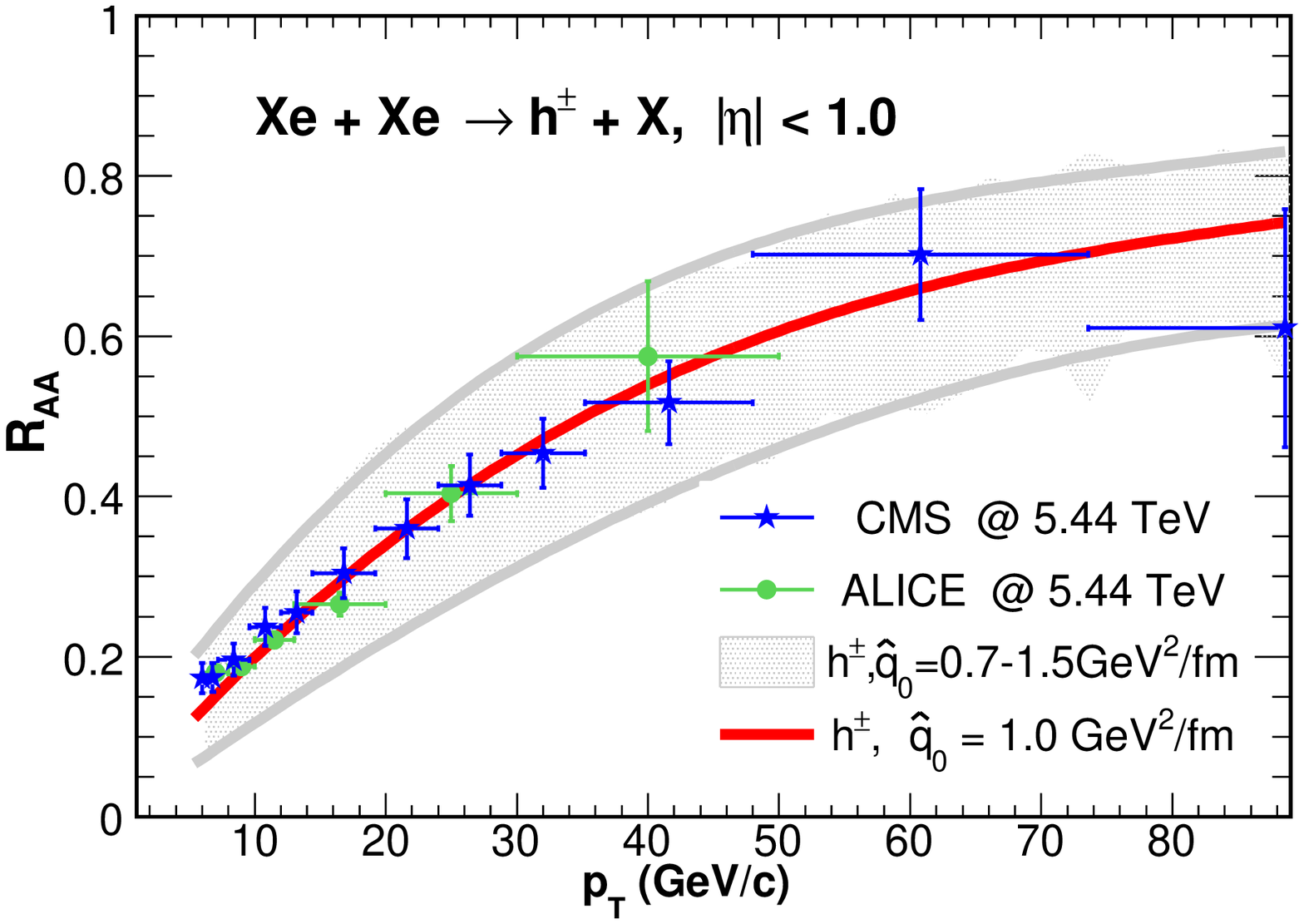}
		    }
	    \end{center}
 	\end{minipage}
 	
 	\begin{minipage}{1\linewidth}
	    \begin{center}
		    \resizebox{0.8\textwidth}{!}{
			    \includegraphics{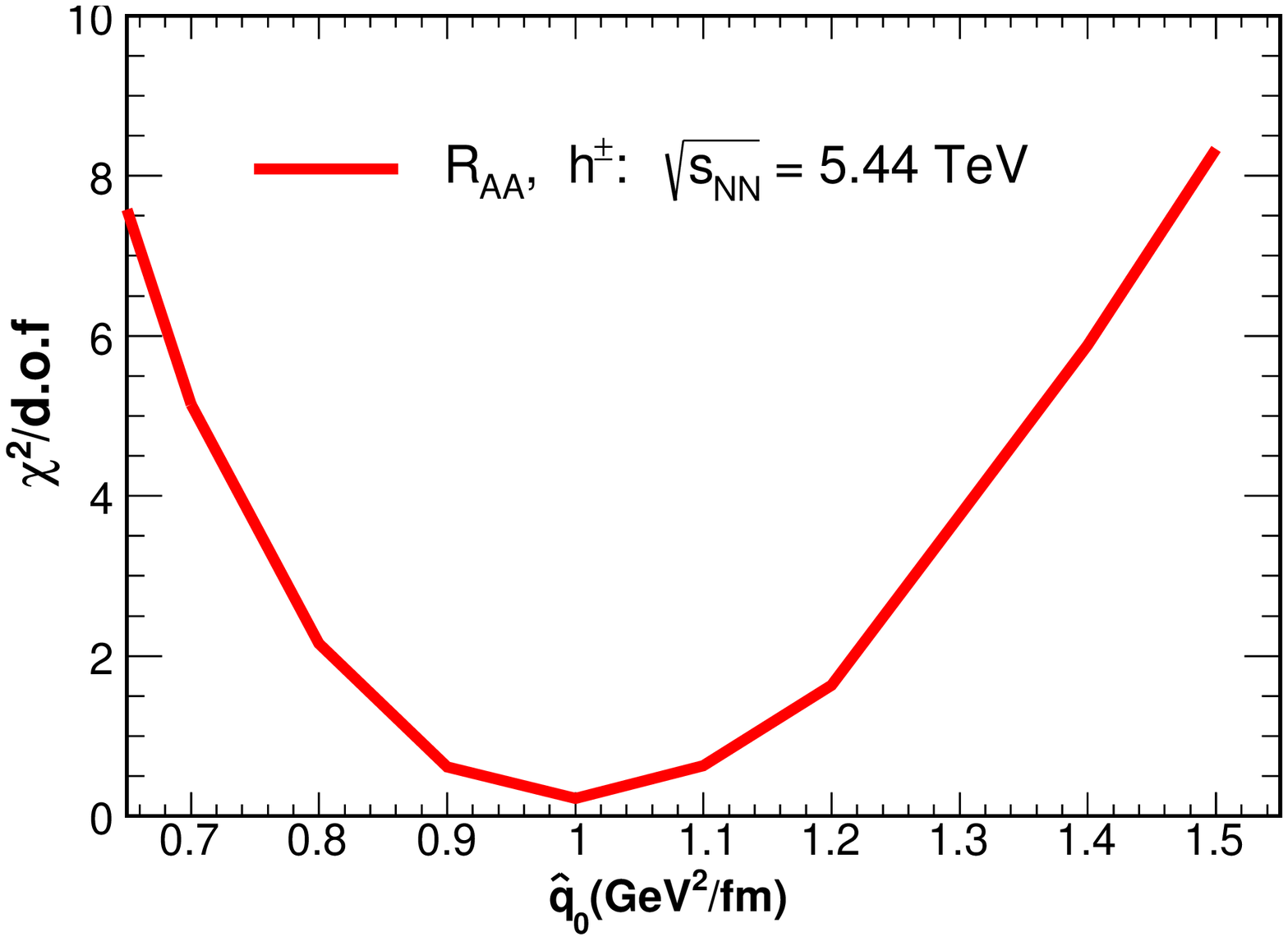}
		    }
	    \end{center}
 	\end{minipage}
	
	\caption{(color online) Upper: Nuclear modification factor $R_{AA}$ of charged hadrons as a function of $p_{\rm T}$ compared with both CMS~\cite{CMS:2018yyx} (blue stars) and ALICE~\cite{ALICE:2018hza} (green dots) data in 0\%-5\% Xe+Xe collisions at $\sqrt{s_{NN}}$ = 5.44~TeV. The best value of $\hat{q}_0 = 1.0 \rm~GeV^2/fm$ is shown by the red solid line. Bottom: $\chi^2/{\rm d.o.f}$ calculations between theoretical results and both CMS~\cite{CMS:2018yyx} and ALICE~\cite{ALICE:2018hza} data of $R_{AA}$ for charged hadrons in Xe+Xe collisions at $\sqrt{s_{NN}}=5.44$~TeV.}
	\label{fig:xexe_h}
\end{figure}
%%%%%%%%%%%%%%%%%%%%%%%%%%%%%%%%%%%%%%%%%%%%%%%%%%%%%

%%%%%%%%%%%%%%%%%%%%%%%%%%%%%%%%%%%%%%%%%%
%%***********    
%%Figure.raaxexe ****
\begin{figure}[htbp]
	\begin{minipage}{1\linewidth}
		\begin{center}
			\resizebox{0.8\textwidth}{!}{
				\includegraphics{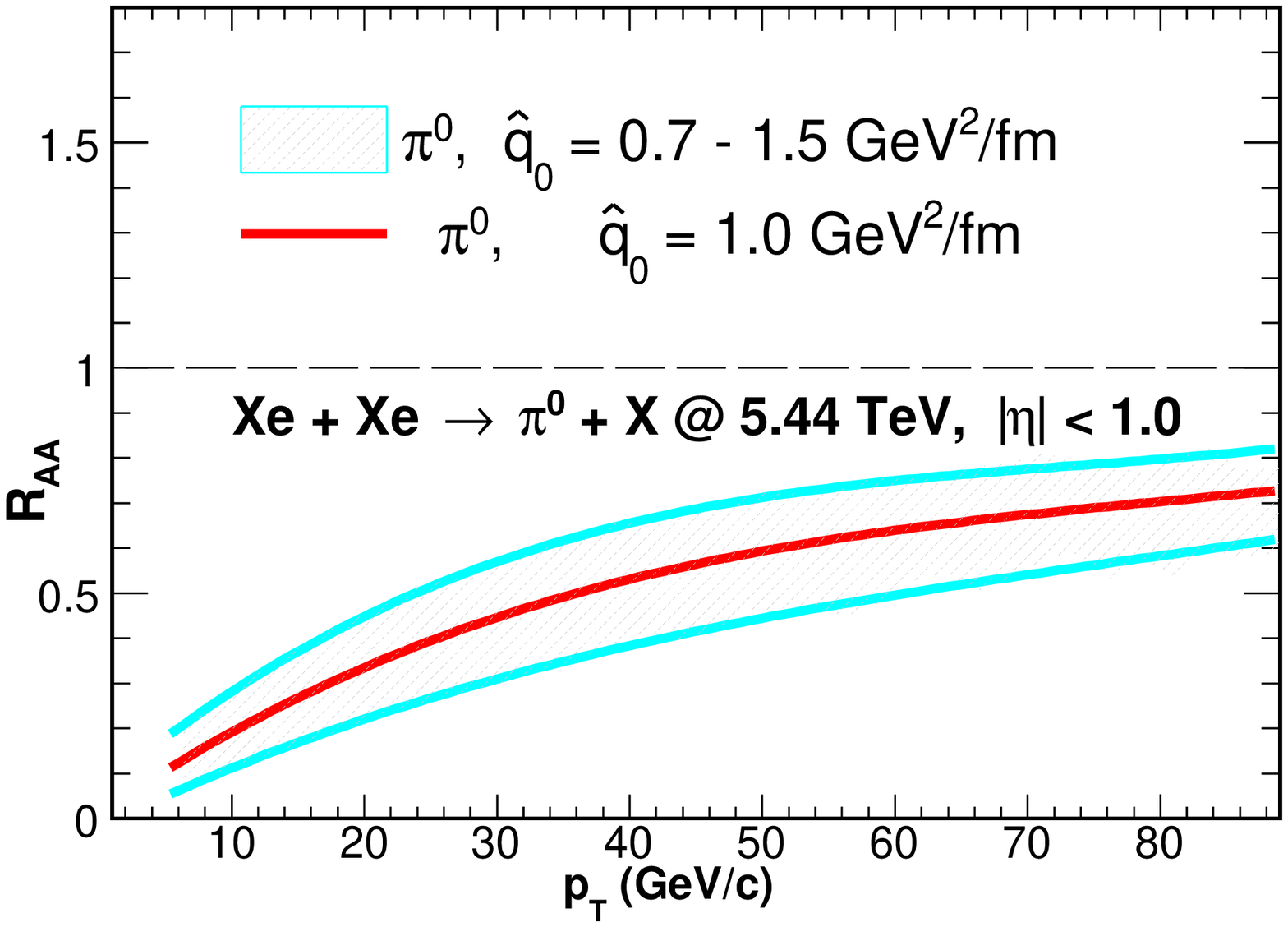}
			}
		\end{center}
	\end{minipage}
	
	\begin{minipage}{1\linewidth}
		\begin{center}
			\resizebox{0.8\textwidth}{!}{
				\includegraphics{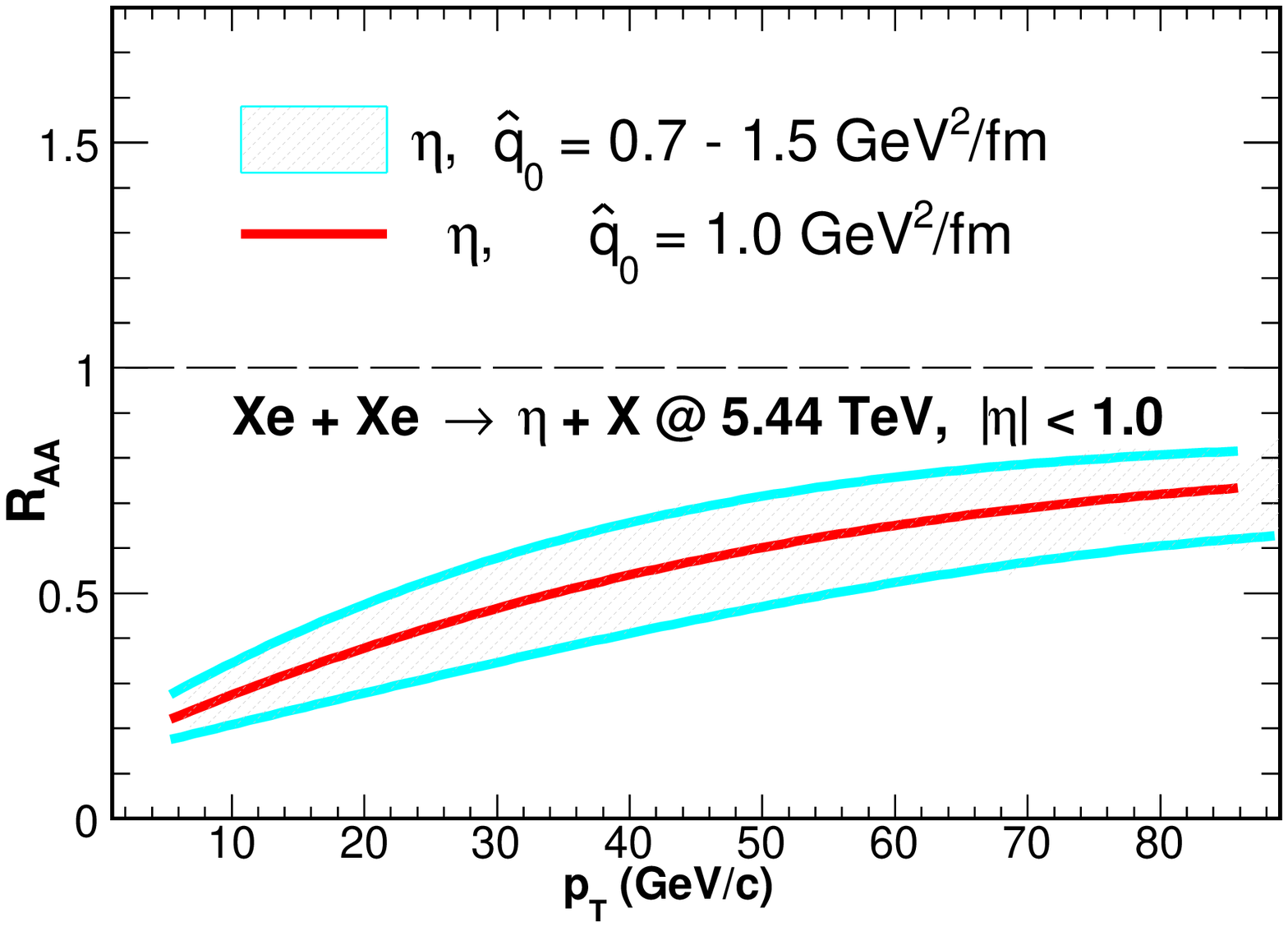}
			}
		\end{center}
	\end{minipage}
	
	\begin{minipage}{1\linewidth}
		\begin{center}
			\resizebox{0.8\textwidth}{!}{
				\includegraphics{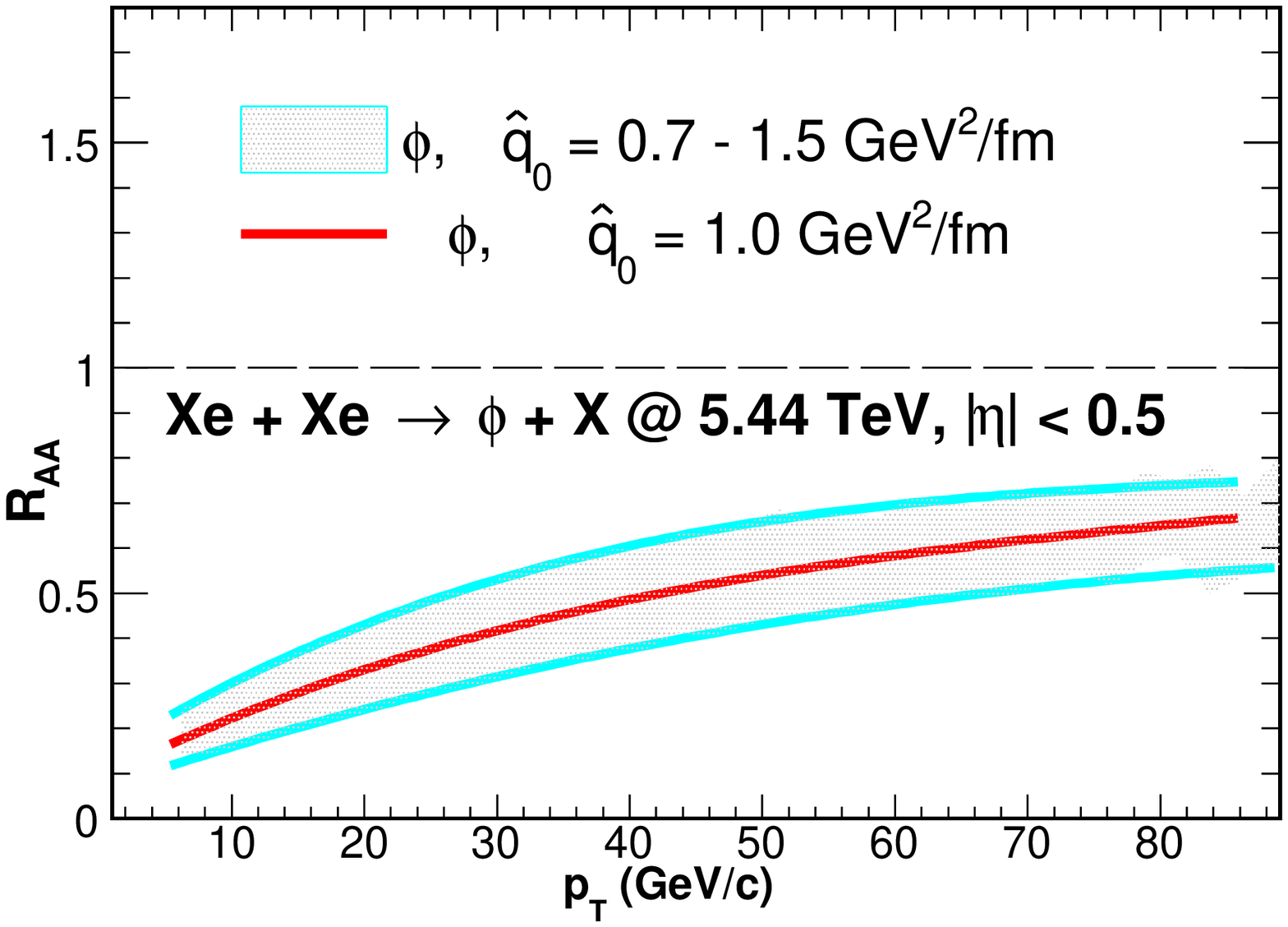}
			}
		\end{center}
	\end{minipage}
	
	\caption{(color online) Nuclear modification factor $R_{AA}$ of $\pi^0$ (upper), $\eta$ (middle), and $\phi$ (bottom) as a function of $p_{\rm T}$ in 0\%-5\% Xe+Xe collisions at $\sqrt{s_{NN}}$ = 5.44~TeV. $\hat{q}_0 = 1.0 \rm~GeV^2/fm$ is represented by the red solid line.}
	\label{fig:raaxexe}
\end{figure}
%%%%%%%%%%%%%%%%%%%%%%%%%%%%%%%%%%%%%%%%%%%%%%%%%%%%%

In the upper panel of Fig.~\ref{fig:xexe_h}, the nuclear modification factor $R_{AA}$ of charged hadrons as a function of $p_{\rm T}$ in the most central (0\%-5\%) Xe+Xe collisions at $\sqrt{s_{NN}}$ = 5.44~TeV is plotted with $\hat{q}_0 = 0.7 - 1.5 \rm~GeV^2/fm$, which is extracted from the $\chi^2/{\rm d.o.f}$ fit in the bottom panel. The theoretical plots of $\hat{q}_0 = 1.0 \rm~GeV^2/fm$ effectively describe both CMS~\cite{CMS:2018yyx} and ALICE~\cite{ALICE:2018hza} data. The bottom panel of Fig.~\ref{fig:xexe_h} shows the $\chi^2/{\rm d.o.f}$ fit of charged hadrons $R_{AA}$ to compare the theoretical results at various values of $\hat{q}_0$ with both CMS~\cite{CMS:2018yyx} and ALICE~\cite{ALICE:2018hza} data in Xe+Xe collisions at $\sqrt{s_{NN}}$ = 5.44~TeV, and the best value of the jet transport coefficient is $\hat{q}_0 = 1.0 \rm~GeV^2/fm$. Within the uncertainty of $\hat{q}_0 = 0.7 - 1.5 \rm~GeV^2/fm$, the theoretical results exhibit a relatively small deviation from experimental data and are thus considered to be in a reasonable range. It is noted that this $\chi^2$ fit is performed at a fixed scale, $\mu$ = 1.0 $p_{\rm T}$, and theoretical uncertainties in nPDFs and FFs as well as from scale variations are not considered. Systematic and statistical uncertainties provided by the LHC are treated equally in our calculations.

%%%%%%%%%%%%%%%%%%%%%%%%%%%%%%%%%%%%%%%%%%
%%***********    
%%Figure.ratio5440 ****
\begin{figure}[htbp]
	
	\begin{minipage}{1\linewidth}
		\begin{center}
			\resizebox{0.8\textwidth}{!}{
				\includegraphics{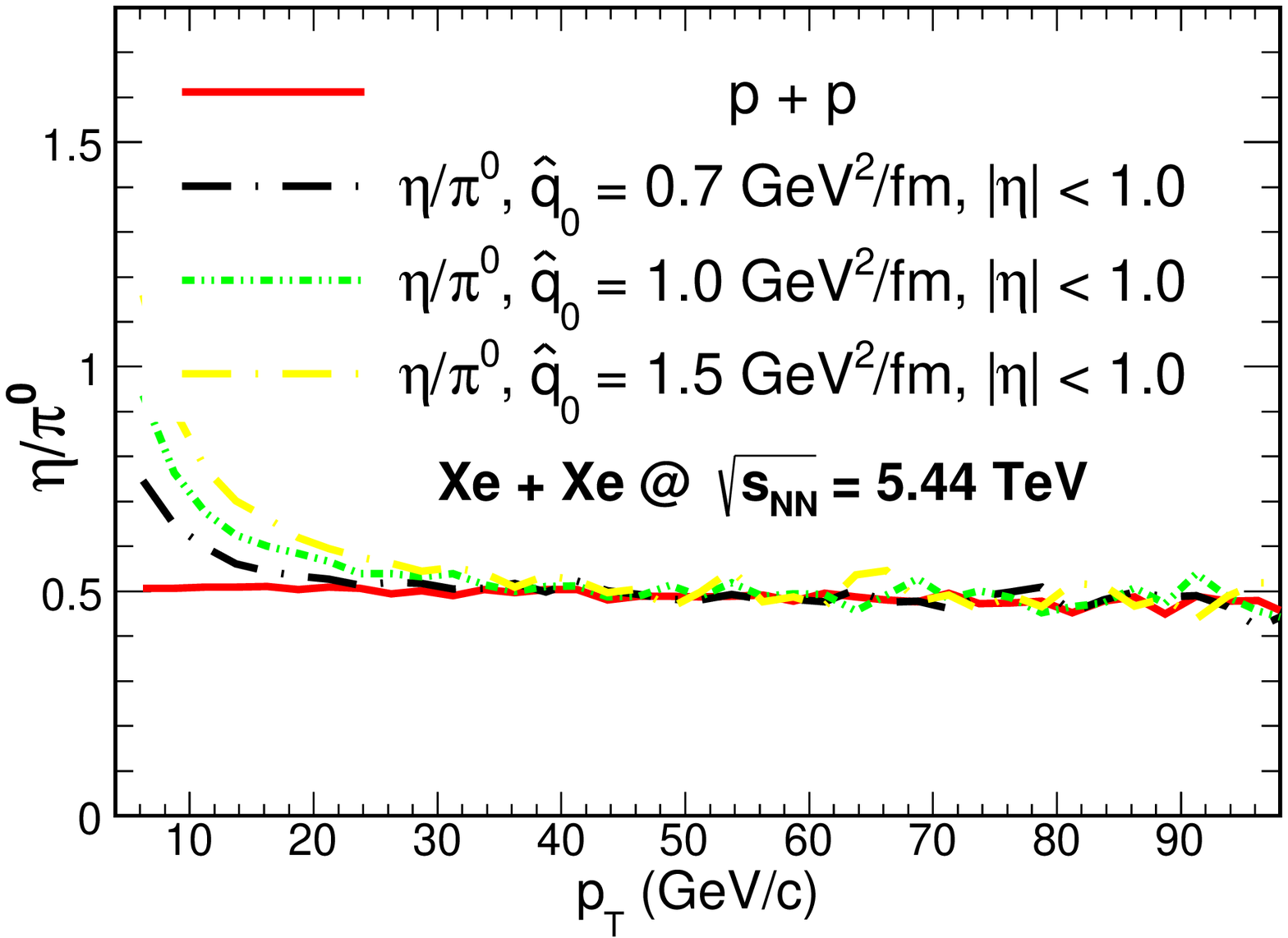}
			}
			%\caption{}
			%\label{}
		\end{center}
	\end{minipage}
	
	\begin{minipage}{1\linewidth}
		\begin{center}
			\resizebox{0.8\textwidth}{!}{
				\includegraphics{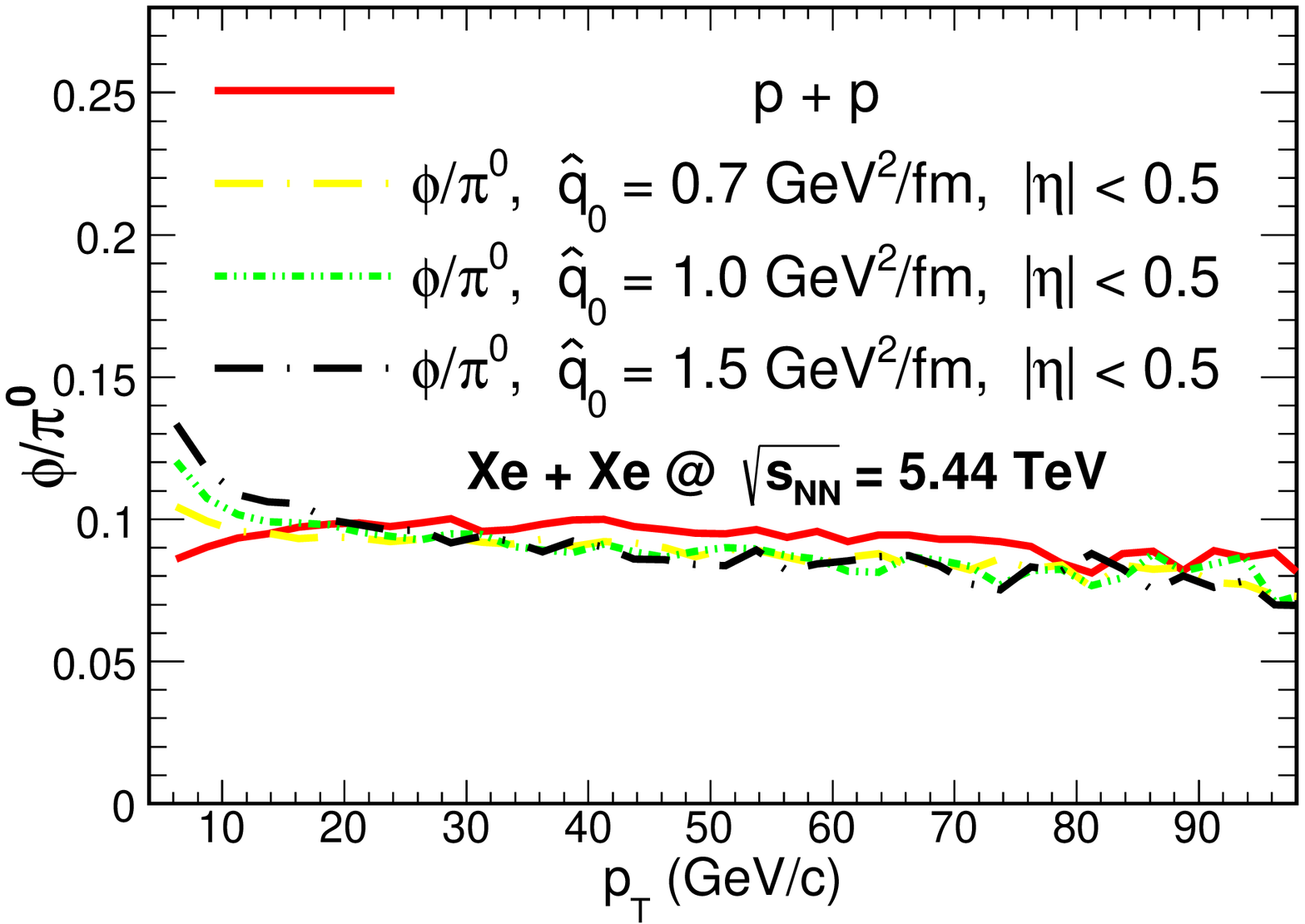}
			}
			%\caption{}
			%\label{}
		\end{center}
	\end{minipage}
	
	\caption{(color online) Particle yield ratios $\eta/\pi^0$ (up) and $\phi/\pi^0$ (down) in 0-5\% Xe+Xe collisions at $\sqrt{s_{NN}}$ = 5.44~TeV with $p$+$p$ reference (red solid line).}
	\label{fig:ratio5440}
\end{figure}
%%%%%%%%%%%%%%%%%%%%%%%%%%%%%%%%%%%%%%%%%%%%%%%%%%%%%

%%%%%%%%%%%%%%%%%%%%%%%%%%%%%%%%%%%%%%%%%%
%%***********    
%%Figure.ratio5020 ****
\begin{figure}[htbp]
	
	\begin{minipage}{1\linewidth}
		\begin{center}
			\resizebox{0.8\textwidth}{!}{
				\includegraphics{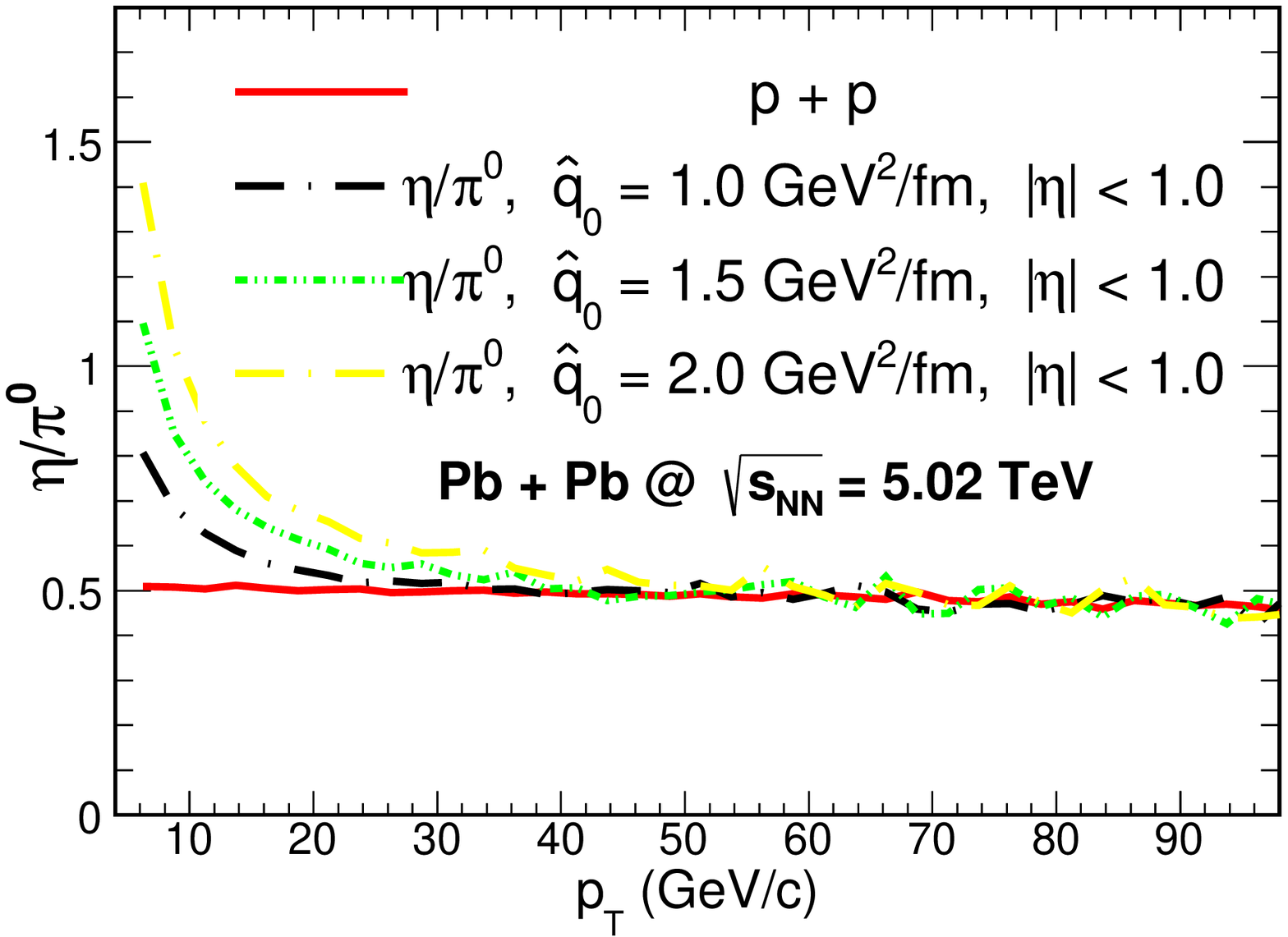}
			}
			%\caption{}
			%\label{}
		\end{center}
	\end{minipage}
	
	\begin{minipage}{1\linewidth}
		\begin{center}
			\resizebox{0.8\textwidth}{!}{
				\includegraphics{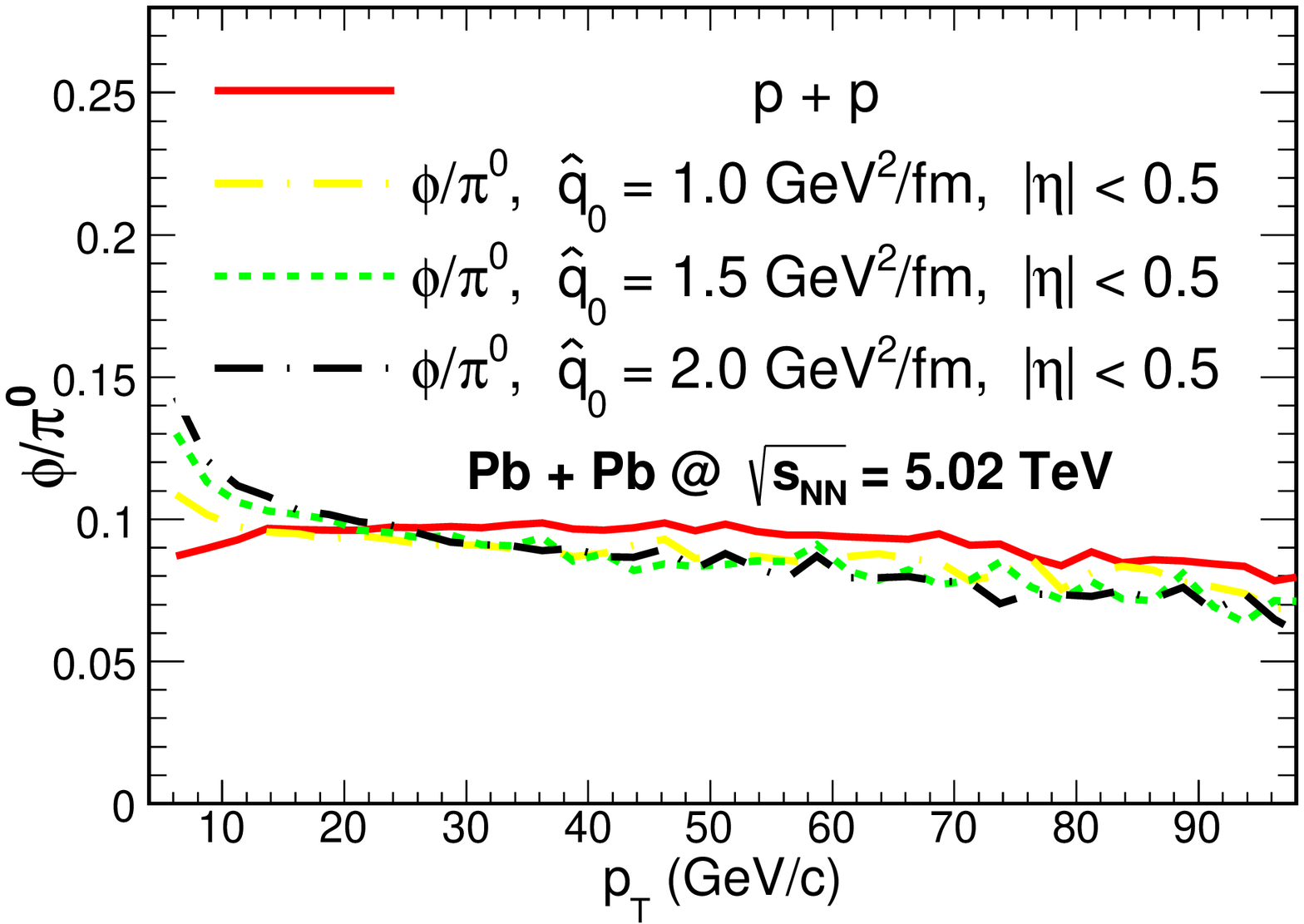}
			}
			%\caption{}
			%\label{}
		\end{center}
	\end{minipage}
	
	\caption{(color online) Particle yield ratios $\eta/\pi^0$ (up) and $\phi/\pi^0$ (down) in 0\%-5\% Pb+Pb collisions at $\sqrt{s_{NN}}$ = 5.02~TeV with $p$+$p$ reference (red solid line).}
	\label{fig:ratio5020}
\end{figure}
%%%%%%%%%%%%%%%%%%%%%%%%%%%%%%%%%%%%%%%%%%%%%%%%%%%%%

Now, we predict the $R_{AA}$ of $\pi^0$ (upper), $\eta$ (middle), and $\phi$ (bottom) as a function of $p_{\rm T}$ in Xe+Xe collisions at $\sqrt{s_{NN}}$ = 5.44~TeV with $\hat{q}_0 = 0.7 - 1.5 \rm~GeV^2/fm$. The results are given in Fig.~\ref{fig:raaxexe}. The trends in the $p_{\rm T}$ dependence of different final-state hadron species are similar, and we can naturally predict the particle ratios in Xe+Xe collisions. In Fig.~\ref{fig:ratio5440}, the particle yield ratios $\eta/\pi^0$ (up) and $\phi/\pi^0$ (down) as functions of $p_{\rm T}$ in Xe+Xe collisions at $\sqrt{s_{NN}}$ = 5.44~TeV with the $p$+$p$ reference (red solid line) are demonstrated. In the upper panel, the $\eta/\pi^0$ ratios in $p$+$p$ collisions are almost independent of $p_{\rm T}$ and remain constant at $\sim$ 0.5, which is exactly the case at 200~GeV and 2.76~TeV calculated in our previous study~\cite{Dai:2015dxa}. 

The particle ratios $\eta/\pi^0$ (up) and $\phi/\pi^0$ (down) as functions of $p_{\rm T}$ in Pb+Pb collisions at $\sqrt{s_{NN}}$ = 5.02~TeV are plotted in Fig.~\ref{fig:ratio5020} as a supplement. We find that $\eta/\pi^0$ in $A$+$A$ collisions coincides with that in $p$+$p$ collisions at larger $p_{\rm T}$, reaching a constant of 0.5 in Au+Au 200~GeV, Pb+Pb 2.76~TeV, Pb+Pb 5.02~TeV, and Xe+Xe 5.44~TeV collisions. This ratio is not affected by different choices of $\hat{q}_0$ at higher $p_{\rm T}$. This is the result of the $\eta$ and $\pi^0$ yields being dominated by quark fragmentation contributions at large $p_{\rm T}$ in $p$+$p$ collisions. The jet quenching effect will enhance the quark contribution fraction so that the ratio in $A$+$A$ collisions remains the same as that in $p$+$p$ collisions, which can be described by the ratio of the corresponding FFs in vacuum~\cite{Dai:2015dxa}.

In the bottom panel of Fig.~\ref{fig:ratio5440} and Fig.~\ref{fig:ratio5020}, the $\phi/\pi^0$ ratio in $p$+$p$ collisions is approximately constant at $\sim$ 0.1 as the final-state $p_{\rm T}$ increases. However, in $A$+$A$ collisions, the ratio $\phi/\pi^0$ slightly decreases with increasing $p_{\rm T}$, and the curves for $A$+$A$ collisions are lower than those for $p$+$p$ collisions. By comprehensively comparing the modifications of $\phi/\pi^0$ induced by the hot and dense medium created at different collision energies from relatively lower collision energies of 200~GeV in Au+Au collisions at the RHIC and 2.76~TeV in Pb+Pb collisions at the LHC~\cite{Dai:2017piq} to higher energies of 5.02~TeV in Pb+Pb collisions and 5.44~TeV in Xe+Xe collisions, the deviation between the two curves in $p$+$p$ and $A$+$A$ collisions is found to decrease with increasing of collision energy in the intermediate and larger region of $p_{\rm T}$. The mechanism behind these deviations is as follows: $\phi$ production at large $p_{\rm T}$ is dominated by the gluon fragmentation contribution, whereas $\pi^0$ production is dominated by quarks in $p$+$p$ collisions. The energy loss effect will depress the gluon contribution fraction and enhance the quark fraction owing to the larger energy loss of gluons than quarks~\cite{Dai:2017piq}.

\section{System size dependence of hadron production suppression}%\romannumeral4.\ 
\label{sec:rxepb}
%%%%%%%%%%%%%%%%%%%%%%%%%%%%%%%%%%%%%%%%%%
%%***********    
%%Figure.raapbpb ****
\begin{figure*}[htbp]
	\begin{minipage}{1\linewidth}
		\begin{center}
			\resizebox{0.75\textwidth}{!}{
				\includegraphics{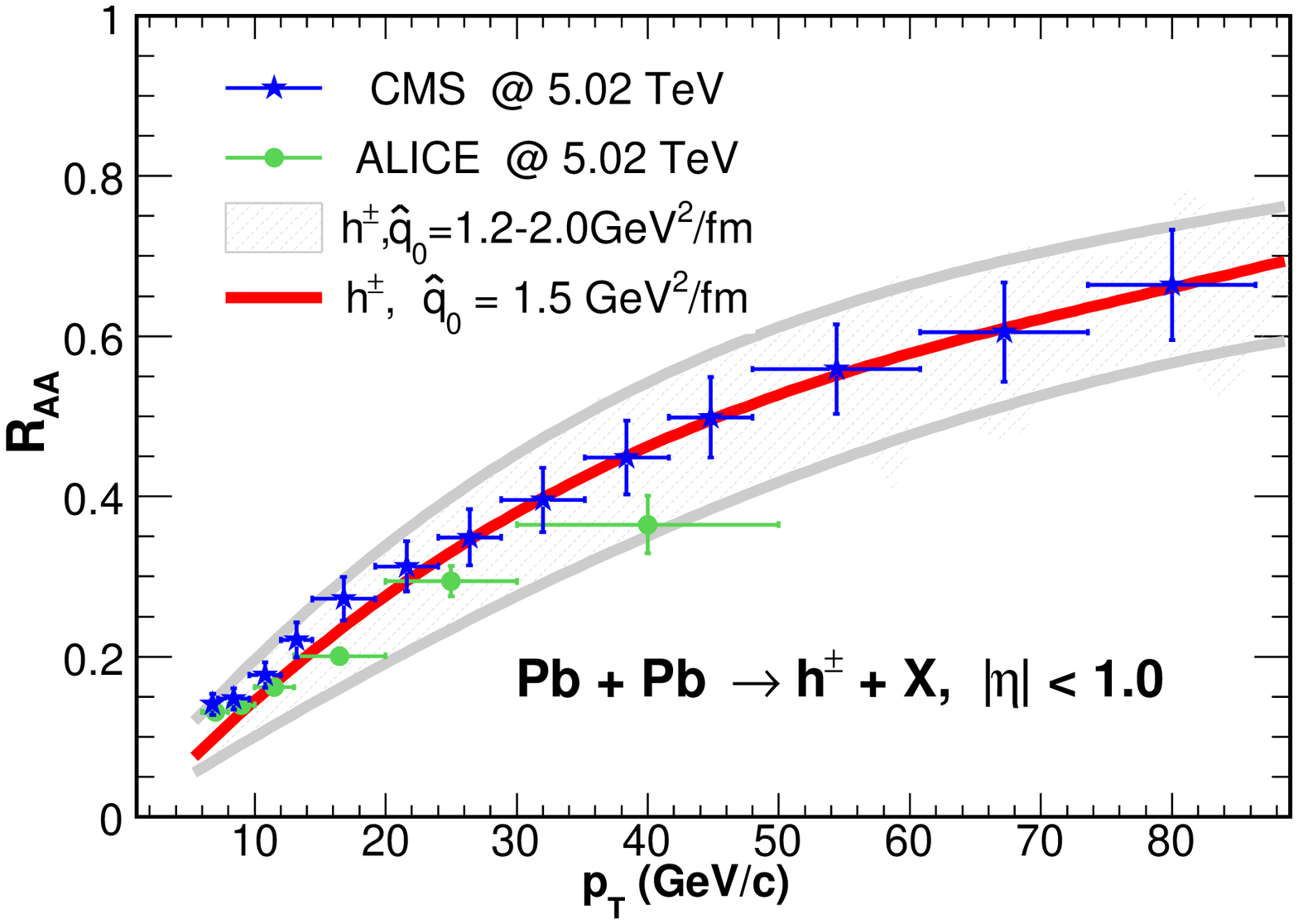}
				\includegraphics{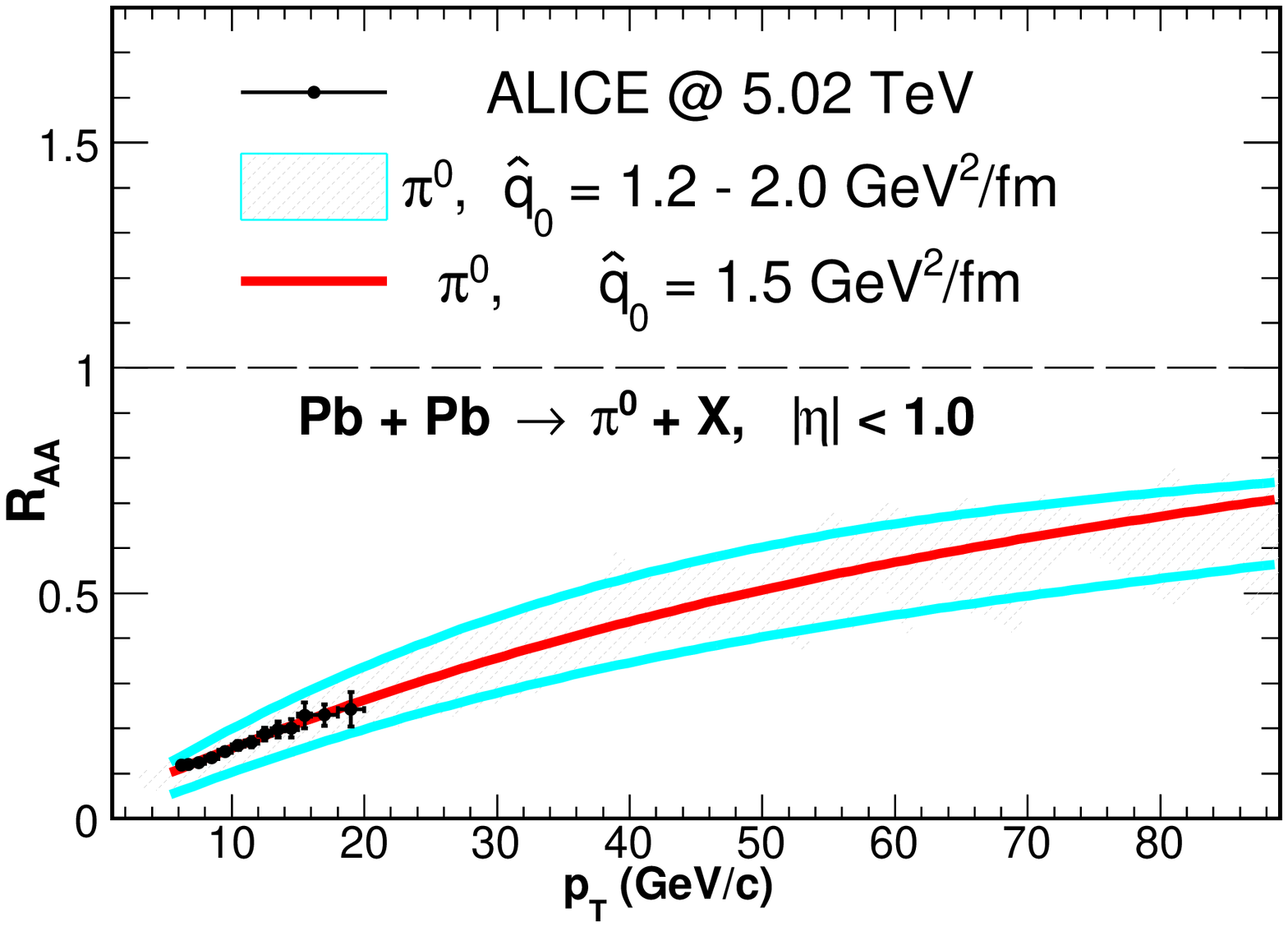}
			}
		\end{center}
	\end{minipage}
	
	\begin{minipage}{1\linewidth}
		\begin{center}
			\resizebox{0.75\textwidth}{!}{
				\includegraphics{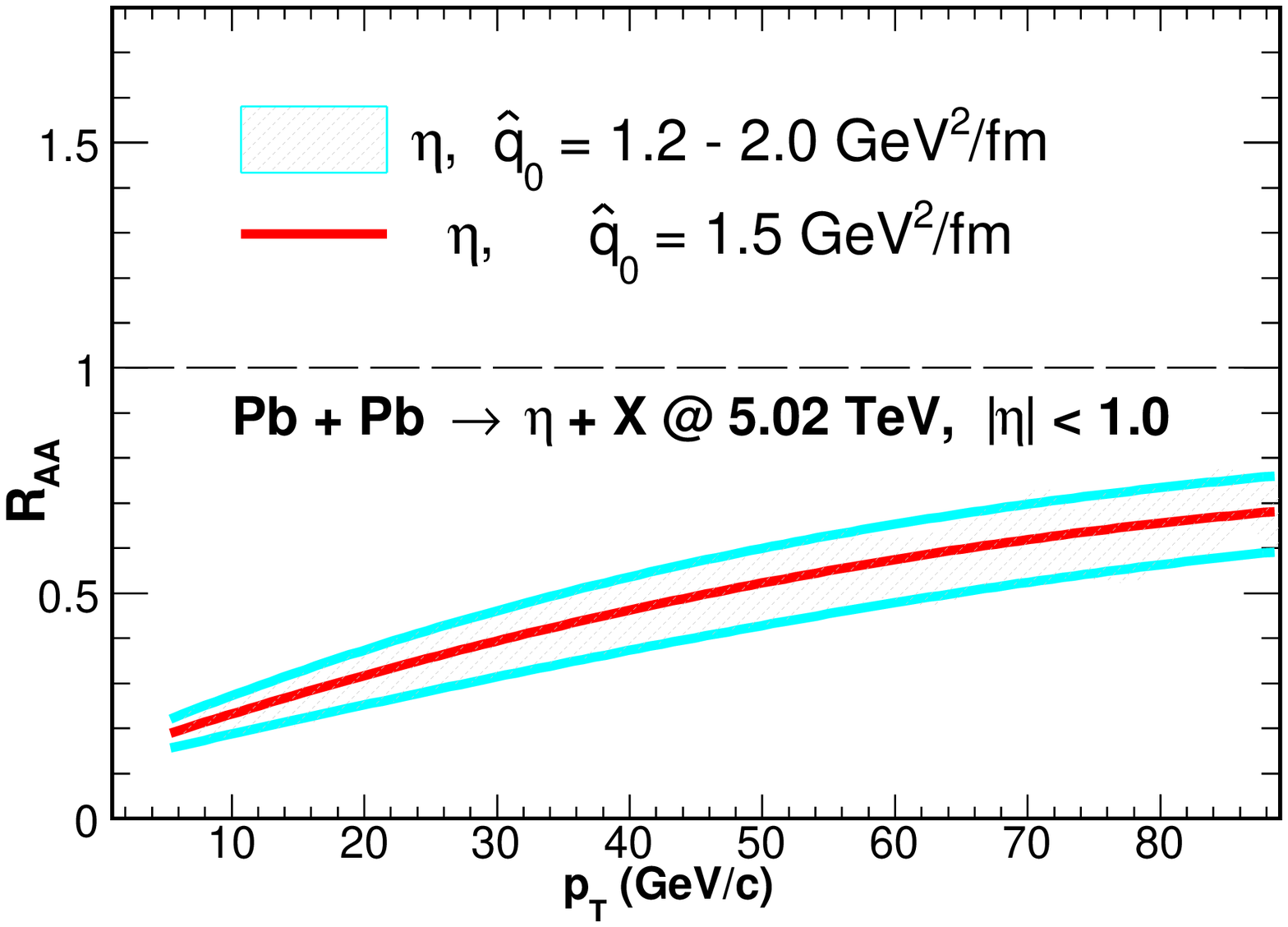}
				\includegraphics{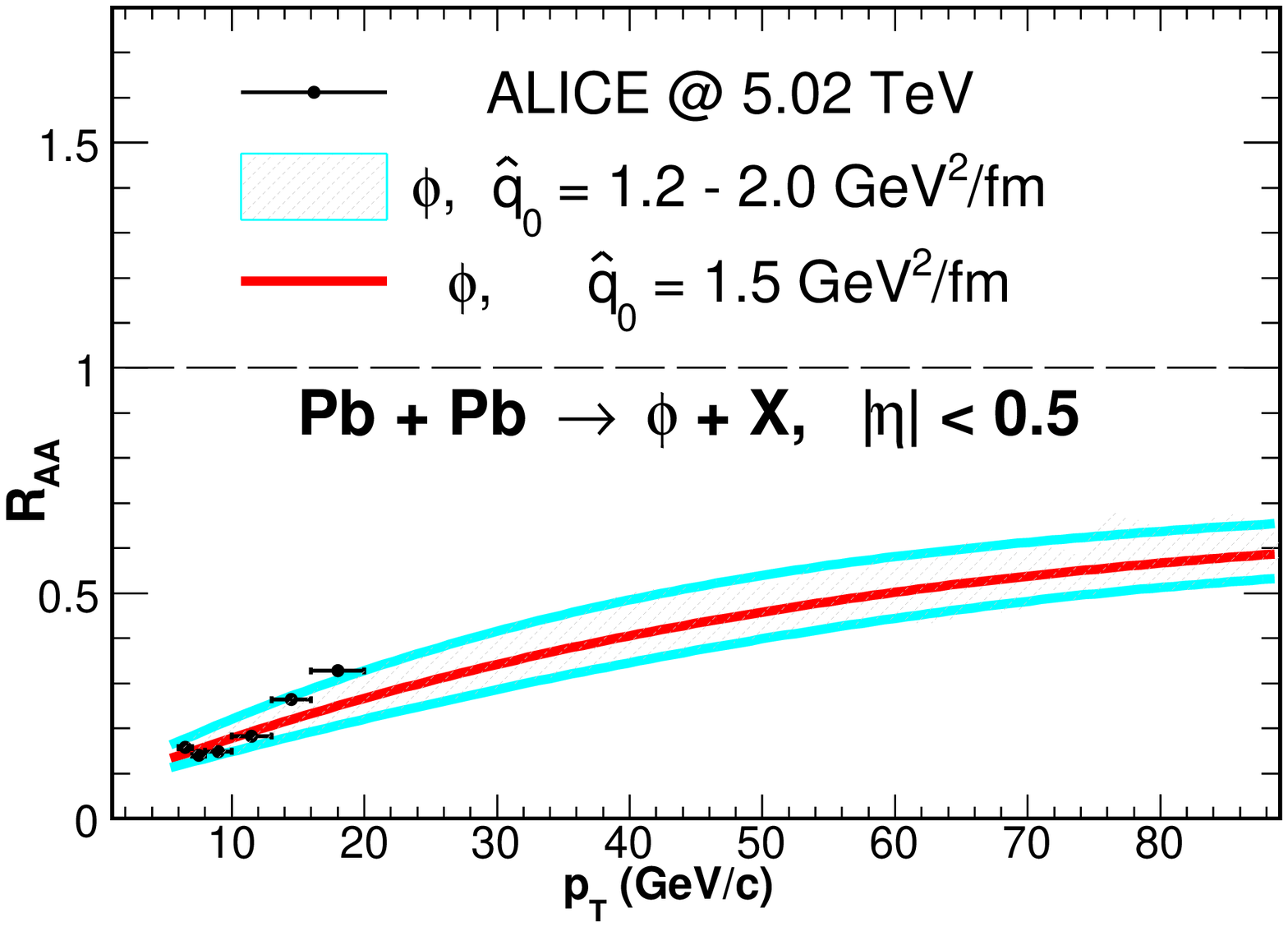}
			}
		\end{center}
	\end{minipage}
	
	\caption{(color online) Nuclear modification factor $R_{AA}$ of charged hadrons (upper left), $\pi^0$ (upper right), $\eta$ (bottom left), and $\phi$ (bottom right) as a function of $p_{\rm T}$ in 0\%-5\% Pb+Pb collisions at $\sqrt{s_{NN}}$ = 5.02~TeV compared with CMS and ALICE data~\cite{CMS:2016xef,ALICE:2018vuu,ALICE:2019hno,ALICE:2021ptz}. $\hat{q}_0 = 1.5 \rm~GeV^2/fm$ is represented by the red solid line.}
	\label{fig:raapbpb}
\end{figure*}
%%%%%%%%%%%%%%%%%%%%%%%%%%%%%%%%%%%%%%%%%%%%%%%%%%%%%

%%%%%%%%%%%%%%%%%%%%%%%%%%%%%%%%%%%%%%%%%%
%%***********    
%%Figure.rxepb ****
\begin{figure}[t]
	\begin{center}
	    \resizebox{0.4\textwidth}{!}{
		    \includegraphics{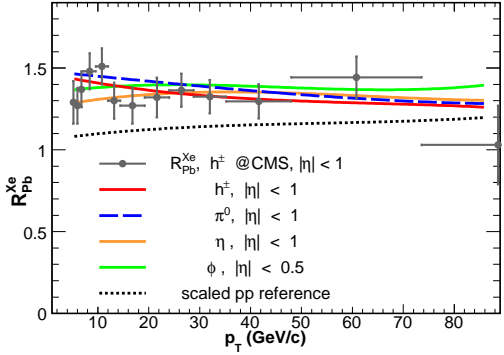}
		}
	\end{center}
	\caption{(color online) Scaled ratio $R^{\rm Xe}_{\rm Pb}$ of charged hadron, $\pi^0,\ \eta$, and $\phi$ production in 0\%-5$\%$ central collisions compared with the CMS data on charged hadrons~\cite{CMS:2018yyx}.}
	\label{fig:rxepb}
\end{figure}
%%%%%%%%%%%%%%%%%%%%%%%%%%%%%%%%%%%%%%%%%%%%%%%%%%%%%

Path-length dependence is a fundamental characteristic of the jet quenching theory, which describes how the energy loss depends on the length of a parton traversing the QCD medium. In October 2017, $^{129}$Xe+$^{129}$Xe collisions at $\sqrt{s_{NN}}$ = 5.44~TeV were measured by the LHC in addition to $^{208}$Pb+$^{208}$Pb collisions. At comparable collision energies (5.02 and 5.44~TeV), the system size was the main difference between Pb+Pb and Xe+Xe collisions. It is interesting to directly compare the medium modifications of inclusive hadron productions in Pb+Pb 5.02~TeV and in Xe+Xe 5.44~TeV systems. Because the $p$+$p$ reference of the nuclear modification factors in both systems can be assumed to be approximately equal, a scaled ratio between the final-state hadron spectra in Xe+Xe and Pb+Pb collisions is proposed~\cite{CMS:2018yyx}.
\begin{eqnarray}
R^{\rm Xe}_{\rm Pb}(p_{\rm T})=\frac{dN^{\rm XeXe}/dp_{\rm T}}{dN^{\rm PbPb}/dp_{\rm T}}\frac{T_{\rm PbPb}}{T_{\rm XeXe}}  \,\,  .
\label{eq:rxepb}
\end{eqnarray}
where $dN^{\rm PbPb,XeXe}/dp_{\rm T}$ are the final-state hadron yields in Pb+Pb (Xe+Xe) collisions, which are scaled by the nuclear overlap function $T_{\rm PbPb, XeXe} = \langle N_{coll}\rangle/\sigma^{pp}$ obtained through the Glauber Model. The total cross sections are $\sigma^{pp}_{5.44} = (68.4 \pm 0.5)$~mb and $\sigma^{pp}_{5.02} = (67.6 \pm 0.6)$~mb. The radii of the $^{207}$Pb and $^{129}$Xe nuclei are $\approx$ 6.62 and 5.36~fm, respectively~\cite{Moller:2015fba}. With comparable collision energies (5.02~TeV and 5.44~TeV) and initial temperatures (502~MeV and 484~MeV, estimated by the (3+1)D viscous hydrodynamic model CLVisc), this observable has been proven by the CMS~\cite{CMS:2018yyx} to be able to experimentally expose the system size (path length) dependence of the jet quenching effect in hadron production in $A$+$A$ collisions. We can see from Eq.~(\ref{eq:rxepb}) that to calculate $R^{\rm Xe}_{\rm Pb}(p_{\rm T})$, final-state leading hadron spectra in Pb+Pb collisions at $\sqrt{s_{NN}}$ = 5.02~TeV are required.

In Fig.~\ref{fig:raapbpb}, we plot $R_{AA}$ distributions as a function of final-state $p_{\rm T}$ for charged hadrons, $\pi^0$, $\eta$, and $\phi$ in Pb+Pb collisions at $\sqrt{s_{NN}}$ = 5.02~TeV compared with LHC data~\cite{CMS:2016xef,ALICE:2018vuu,ALICE:2019hno,ALICE:2021ptz}. In the upper left plots, we present the case of charged hadrons; the theoretical plots within the uncertainty of $\hat{q}_0 = 1.2 - 2.0 \rm~GeV^2/fm$ can describe both ALICE~\cite{ALICE:2018vuu} and CMS~\cite{CMS:2016xef} data well, with a best value of $\hat{q}_0 = 1.5 \rm~GeV^2/fm$. The same theoretical uncertainty in $\hat{q}_0$ is utilized to describe the $R_{AA}$ distributions of $\pi^0,\ \eta$, and $\phi$ in the other plots of Fig.~\ref{fig:raapbpb}, where we find that the results of $\pi^0$ and $\phi$ can describe ALICE data~\cite{ALICE:2019hno,ALICE:2021ptz}. This value is higher than $\hat{q}_0 = 1.0 \rm~GeV^2/fm$ in Xe+Xe collisions at $\sqrt{s_{NN}}$ = 5.44~TeV, corresponding $\hat{q}/T^3$ range with an initial temperature $T_0$ = 502~MeV at initial time $\tau_0 = 0.6$ fm/$c$, and the central position is 1.6 $\sim$ 2.8 in Pb+Pb collisions at $\sqrt{s_{NN}}=5.02$~TeV. This value in Xe+Xe collisions at $\sqrt{s_{NN}}=5.44$~TeV is 1.8 $\sim$ 2.6 with an initial temperature $T_0$ = 484~MeV. We note that the JET Collaboration found that $\hat q/T^3$ has a dependence on the medium temperature $T$~\cite{JET:2013cls}, and the baryon chemical potential $\mu_B$ dependence of $\hat{q}$ as well as jet properties such as the momentum of jet partons, mass, flavor, and the strong coupling constant is investigated in Refs.~\cite{Moreau:2019vhw,Grishmanovskii:2022tpb}. Therefore, at different collision energies, initial temperatures, and nucleus radii, the extracted best value of $\hat{q}_0$ is expected to differ. More experimental data will help us better constrain such an extraction, especially in Xe+Xe collisions.

By comparing the results in Fig.~\ref{fig:raaxexe} and Fig.~\ref{fig:raapbpb}, we find that the $R_{AA}$ distributions of all four final-state hadrons in Xe+Xe 5.44~TeV collisions are slightly less suppressed than those in Pb+Pb 5.02~TeV collisions, which agrees with the experimental observation of charged hadron $R_{AA}$ measurements by ALICE~\cite{ALICE:2018hza} and CMS~\cite{CMS:2018yyx}.

Now, we are able to demonstrate the scaled ratios $R^{\rm Xe}_{\rm Pb}$ of charged hadrons, $\pi^0,\ \eta$, and $\phi$ as functions of $p_{\rm T}$ in 0\%-5$\%$, the most central collisions in Fig.~\ref{fig:rxepb}, with the experimental data points of charged hadron production provided by CMS~\cite{CMS:2018yyx}, and the ratio of the $p$+$p$ reference at 5.44~TeV over that at 5.02~TeV, $\left(\frac{d\sigma_{5.44}^{pp}}{dp_{\rm T}}/\frac{d\sigma_{5.02}^{pp}}{dp_{\rm T}}\right)$, is also plotted as the ``scaled $p$+$p$ reference" (dotted line), which represents contributions from the differences in collision energies. The scaled ratios of all four final-state hadrons are found at the best value of $\hat{q}_0$ in both the Xe+Xe and Pb+Pb systems. The calculation results for charged hadrons agree with the experimental data within the margin of error. The predictions for $\pi^0,\ \eta$, and $\phi$ almost coincide with the charged hadron curve. The deviation of $R^{\rm Xe}_{\rm Pb}$ from the $p$+$p$ reference ratio suggests that production suppression in Xe+Xe collisions is smaller than that in Pb+Pb collisions and shows that a larger system size contributes more suppression to the nuclear modification factor, which is consistent with the comparison of $R_{AA}$ in these two systems. We note that although cold nuclear matter (CNM) effects from nPDFs may impact particle yields at large $p_{\rm T}$ in $A$+$A$ collisions, their contributions to $R^{\rm Xe}_{\rm Pb}$ are negligible with the understanding that CNM effects largely cancel each other out where the scaled ratio is concerned.

\section{Summary}%\romannumeral5.\ 
\label{sec:summary}
In this study, we predict $\pi^0$, $\eta$, and $\phi$ yields in Xe+Xe collisions at $\sqrt{s_{NN}}$ = 5.44~TeV within the NLO pQCD improved parton model by considering the jet quenching effect with the higher-twist approach. The jet transport coefficient $\hat{q}_0$ is extracted by fitting the $R_{AA}$ of charged hadron production with both ALICE and CMS data. The nuclear modification factors of $\pi^0$, $\eta$, and $\phi$ as functions of $p_{\rm T}$ are then predicted. Cases involving Pb+Pb collisions at $\sqrt{s_{NN}}$ = 5.02~TeV are also provided for comparison. In all systems, we study from Au+Au at 200~GeV to Pb+Pb at 2.76~TeV and 5.02~TeV to Xe+Xe at 5.44~TeV. The curves of $\eta/\pi^0$ in $p$+$p$ and $A$+$A$ collisions coincide at a constant of 0.5 and show little dependence on $p_{\rm T}$; however, the $A$+$A$ curves of $\phi/\pi^0$ consistently deviate from their $p$+$p$ references, and only the deviations exhibit a slight dependence on the collision energies. The nuclear modification factors of $\pi^0$, $\eta$, $\phi$, and charged hadrons in Xe+Xe 5.44~TeV collisions are slightly less suppressed than those in Pb+Pb 5.02~TeV collisions. The theoretical results of the scaled ratio $R^{\rm Xe}_{\rm Pb}$ of final-state $\pi^0$, $\eta$, and $\phi$ coincide with the curve of charged hadron production, which can describe the CMS data within the margin of error, indicating that the path-length effect is independent of the species of final-state hadrons.

\section*{ACKNOWLEDGEMENTS}
We express our thanks to Dr. Xiang-Yu Wu for providing detailed profiles of the (3+1)D viscous hydrodynamic model CLVisc and Dr. Man Xie and Guo-Yang Ma for helpful discussions. This research is supported by the Guangdong Major Project of Basic and Applied Basic Research No. 2020B0301030008, the Natural Science Foundation of China with Project Nos. 11935007 and 11805167.


\begin{thebibliography}{99}

%\cite{Wang:1991xy}
\bibitem{Wang:1991xy}
X.~N.~Wang and M.~Gyulassy,
%``Gluon shadowing and jet quenching in A + A collisions at s**(1/2) = 200-GeV,''
Phys.\ Rev.\ Lett.\  {\bf 68}, 1480 (1992).
%  doi:10.1103/PhysRevLett.68.1480
%%CITATION = doi:10.1103/PhysRevLett.68.1480;%%

%\cite{Gyulassy:2003mc}
\bibitem{Gyulassy:2003mc}
M.~Gyulassy, I.~Vitev, X.~N.~Wang and B.~W.~Zhang,
%``Jet quenching and radiative energy loss in dense nuclear matter,''
In *Hwa, R.C. (ed.) et al.: Quark gluon plasma* 123-191
%doi:10.1142/97898127955330003
[nucl-th/0302077].
%%CITATION = doi:10.1142/9789812795533_0003;%%

%\cite{Qin:2015srf}
\bibitem{Qin:2015srf}
G.~Y.~Qin and X.~N.~Wang,
%``Jet quenching in high-energy heavy-ion collisions,''
Int. J. Mod. Phys. E \textbf{24} (2015) no.11, 1530014
doi:10.1142/S0218301315300143
[arXiv:1511.00790 [hep-ph]].
%221 citations counted in INSPIRE as of 20 Mar 2022

%\cite{Ma:2010dv}
\bibitem{Ma:2010dv}
G.~L.~Ma and X.~N.~Wang,
%``Jets, Mach cone, hot spots, ridges, harmonic flow, dihadron and $\gamma$-hadron correlation in high-energy heavy-ion collisions,''
Phys. Rev. Lett. \textbf{106} (2011), 162301
doi:10.1103/PhysRevLett.106.162301
[arXiv:1011.5249 [nucl-th]].
%114 citations counted in INSPIRE as of 20 Mar 2022

%\cite{Fochler:2011en}
\bibitem{Fochler:2011en}
O.~Fochler, J.~Uphoff, Z.~Xu and C.~Greiner,
%``Jet quenching and elliptic flow at RHIC and LHC within a pQCD-based partonic transport model,''
J. Phys. G \textbf{38} (2011), 124152
doi:10.1088/0954-3899/38/12/124152
[arXiv:1107.0130 [hep-ph]].
%67 citations counted in INSPIRE as of 20 Mar 2022

%\cite{Zhang:2003wk}
\bibitem{Zhang:2003wk}
B.~Zhang, E.~Wang and X.~Wang,
%``Heavy quark energy loss in nuclear medium,''
Phys. Rev. Lett. \textbf{93} (2004), 072301
doi:10.1103/PhysRevLett.93.072301
[arXiv:nucl-th/0309040 [nucl-th]].

%\cite{Guo:2000nz}
\bibitem{Guo:2000nz}
X.~f.~Guo and X.~N.~Wang,
%``Multiple scattering, parton energy loss and modified fragmentation functions in deeply inelastic e A scattering,''
Phys. Rev. Lett. \textbf{85} (2000), 3591-3594
doi:10.1103/PhysRevLett.85.3591
[arXiv:hep-ph/0005044 [hep-ph]].

%\cite{Chen:2010te}
\bibitem{Chen:2010te}
X.~Chen, C.~Greiner, E.~Wang, X.~N.~Wang and Z.~Xu,
%``Bulk matter evolution and extraction of jet transport parameter in heavy-ion collisions at RHIC,''
Phys.\ Rev.\ C {\bf 81}, 064908 (2010).
%  doi:10.1103/PhysRevC.81.064908
%  [arXiv:1002.1165 [nucl-th]].
%%CITATION = doi:10.1103/PhysRevC.81.064908;%%

%\cite{Chen:2011vt}
\bibitem{Chen:2011vt}
X.~Chen, T.~Hirano, E.~Wang, X.~N.~Wang and H.~Zhang,
%``Suppression of high $p_{T}$ hadrons in $Pb+Pb$ Collisions at LHC,''
Phys.\ Rev.\ C {\bf 84}, 034902 (2011).
%  doi:10.1103/PhysRevC.84.034902
%  [arXiv:1102.5614 [nucl-th]].
%%CITATION = doi:10.1103/PhysRevC.84.034902;%%

%\cite{Zhang:2003yn}
\bibitem{Zhang:2003yn}
B.~W.~Zhang and X.~N.~Wang,
%``Multiple parton scattering in nuclei: Beyond helicity amplitude approximation,''
Nucl.\ Phys.\ A {\bf 720}, 429 (2003)
%doi:10.1016/S0375-9474(03)01003-0
[hep-ph/0301195].
%%CITATION = doi:10.1016/S0375-9474(03)01003-0;%%

%\cite{Schafer:2007xh}
\bibitem{Schafer:2007xh}
A.~Schafer, X.~N.~Wang and B.~W.~Zhang,
%``Multiple Parton Scattering in Nuclei: Quark-quark Scattering,''
Nucl.\ Phys.\ A {\bf 793}, 128 (2007)
%  doi:10.1016/j.nuclphysa.2007.06.009
[arXiv:0704.0106 [hep-ph]].
%%CITATION = doi:10.1016/j.nuclphysa.2007.06.009;%%

%\cite{Liu:2015vna}
\bibitem{Liu:2015vna}
Z.~Q.~Liu, H.~Zhang, B.~W.~Zhang and E.~Wang,
%``Quantifying jet transport properties via large $p_T$ hadron production,''
Eur. Phys. J. C \textbf{76} (2016) no.1, 20
doi:10.1140/epjc/s10052-016-3885-3
[arXiv:1506.02840 [nucl-th]].

%\cite{Kidonakis:2000gi}
\bibitem{Kidonakis:2000gi}
N.~Kidonakis and J.~F.~Owens,
%``Effects of higher order threshold corrections in high E(T) jet production,''
Phys. Rev. D \textbf{63} (2001), 054019
doi:10.1103/PhysRevD.63.054019
[arXiv:hep-ph/0007268 [hep-ph]].

%%%%%%%%%%%%%%%%%%%%%%  WD %%%%%%%%%%%%%%%%%%%%%%%%%%%%%
%\cite{Dai:2015dxa}
\bibitem{Dai:2015dxa}
W.~Dai, X.~Chen, B.~Zhang and E.~Wang,
%``$\eta$ meson production of high-energy nuclear collisions at NLO,''
Phys. Lett. B \textbf{750} (2015), 390-395
doi:10.1016/j.physletb.2015.09.053
[arXiv:1506.00838 [nucl-th]].

%\cite{Dai:2017tuy}
\bibitem{Dai:2017tuy}
W.~Dai, B.~W.~Zhang and E.~Wang,
%``Production of $\rho^{0}$ meson with large $p_T$ at NLO in heavy-ion collisions,''
Phys.\ Rev.\ C {\bf 98} (2018) 024901
doi:10.1103/PhysRevC.98.024901
[arXiv:1701.04147 [nucl-th]].
%%CITATION = doi:10.1103/PhysRevC.98.024901;%%

%\cite{Dai:2017piq}
\bibitem{Dai:2017piq}
W.~Dai, X.~F.~Chen, B.~W.~Zhang, H.~Z.~Zhang and E.~Wang,
%``Nuclear suppression of the $\phi $ meson yields with large $p_T$ at the RHIC and the LHC,''
Eur.\ Phys.\ J.\ C {\bf 77} (2017) no.8,  571
doi:10.1140/epjc/s10052-017-5136-7
[arXiv:1702.01614 [nucl-th]].
%%CITATION = doi:10.1140/epjc/s10052-017-5136-7;%%

%\cite{Ma:2018swx}
\bibitem{Ma:2018swx}
G.~Y.~Ma, W.~Dai, B.~W.~Zhang and E.~K.~Wang,
%``NLO Productions of $\omega $ and $K^0_{\mathrm{S}}$ with a global extraction of the jet transport parameter in heavy-ion collisions,''
Eur.\ Phys.\ J.\ C {\bf 79} (2019) no.6,  518
doi:10.1140/epjc/s10052-019-7005-z
[arXiv:1812.02033 [nucl-th]].
%%CITATION = doi:10.1140/epjc/s10052-019-7005-z;%%
%%%%%%%%%%%%%%%%%%%%%%  WD

%%%%%%%%%%%%%%%%%%%%%%  200 GeV %%%%%%%%%%%%%%%%%%%%%%%%%%%%%
%\cite{PHENIX:2006ujp}
\bibitem{PHENIX:2006ujp}
S.~S.~Adler \textit{et al.} [PHENIX],
%``Common suppression pattern of eta and pi0 mesons at high transverse momentum in Au+Au collisions at S(NN)**(1/2) = 200-GeV,''
Phys. Rev. Lett. \textbf{96} (2006), 202301
doi:10.1103/PhysRevLett.96.202301
[arXiv:nucl-ex/0601037 [nucl-ex]].

%\cite{PHENIX:2010hvs}
\bibitem{PHENIX:2010hvs}
A.~Adare \textit{et al.} [PHENIX],
%``Cross section and double helicity asymmetry for $\eta$ mesons and their comparison to neutral pion production in $p+p$ collisions at $\sqrt{s}$=200 GeV,''
Phys. Rev. D \textbf{83} (2011), 032001
doi:10.1103/PhysRevD.83.032001
[arXiv:1009.6224 [hep-ex]].

%\cite{STAR:2011iap}
\bibitem{STAR:2011iap}
G.~Agakishiev \textit{et al.} [STAR],
%``Identified hadron compositions in p+p and Au+Au collisions at high transverse momenta at $\sqrt{s_{_{NN}}} = 200$ GeV,''
Phys. Rev. Lett. \textbf{108} (2012), 072302
doi:10.1103/PhysRevLett.108.072302
[arXiv:1110.0579 [nucl-ex]].
%%%%%%%%%%%%%%%%%%%%%%  200 GeV

%%%%%%%%%%%%%%%%%%%%%%  qhat %%%%%%%%%%%%%%%%%%%%%%%%%%%%%
%\cite{JET:2013cls}
\bibitem{JET:2013cls}
K.~M.~Burke \textit{et al.} [JET],
%``Extracting the jet transport coefficient from jet quenching in high-energy heavy-ion collisions,''
Phys. Rev. C \textbf{90} (2014) no.1, 014909
doi:10.1103/PhysRevC.90.014909
[arXiv:1312.5003 [nucl-th]].

%\cite{Xie:2019oxg}
\bibitem{Xie:2019oxg}
M.~Xie, S.~Y.~Wei, G.~Y.~Qin and H.~Z.~Zhang,
%``Extracting jet transport coefficient via single hadron and dihadron productions in high-energy heavy-ion collisions,''
Eur. Phys. J. C \textbf{79} (2019) no.7, 589
doi:10.1140/epjc/s10052-019-7100-1
[arXiv:1901.04155 [hep-ph]].

%%%%%%%%%%%%%%%%%%%%%%%%%%%%% Xe + Xe %%%%%%%%%%%%%%%%%%%%%%%%%%%%%%%%%%%%%%
%\cite{CMS:2018yyx}
\bibitem{CMS:2018yyx}
A.~M.~Sirunyan \textit{et al.} [CMS],
%``Charged-particle nuclear modification factors in XeXe collisions at $ \sqrt{s_{\mathrm{NN}}} = 5.44 $~TeV,''
JHEP \textbf{10} (2018), 138
doi:10.1007/JHEP10(2018)138
[arXiv:1809.00201 [hep-ex]].

%\cite{ALICE:2018hza}
\bibitem{ALICE:2018hza}
S.~Acharya \textit{et al.} [ALICE],
%``Transverse momentum spectra and nuclear modification factors of charged particles in Xe-Xe collisions at $\sqrt{s_{\rm NN}}$ = 5.44~TeV,''
Phys. Lett. B \textbf{788} (2019), 166-179
doi:10.1016/j.physletb.2018.10.052
[arXiv:1805.04399 [nucl-ex]].
%%%%%%%%%%%%%%%%%%%%%%%%%%%%% Xe + Xe

%\cite{Dulat:2015mca}
\bibitem{Dulat:2015mca}
S.~Dulat, T.~J.~Hou, J.~Gao, M.~Guzzi, J.~Huston, P.~Nadolsky, J.~Pumplin, C.~Schmidt, D.~Stump and C.~P.~Yuan,
%``New parton distribution functions from a global analysis of quantum chromodynamics,''
Phys. Rev. D \textbf{93} (2016) no.3, 033006
doi:10.1103/PhysRevD.93.033006
[arXiv:1506.07443 [hep-ph]].

%\cite{Kniehl:2000fe}
\bibitem{Kniehl:2000fe}
B.~A.~Kniehl, G.~Kramer and B.~Potter,
%``Fragmentation functions for pions, kaons, and protons at next-to-leading order,''
Nucl. Phys. B \textbf{582} (2000), 514-536
doi:10.1016/S0550-3213(00)00303-5
[arXiv:hep-ph/0010289 [hep-ph]].

%\cite{Aidala:2010bn}
\bibitem{Aidala:2010bn}
C.~A.~Aidala, F.~Ellinghaus, R.~Sassot, J.~P.~Seele and M.~Stratmann,
%``Global Analysis of Fragmentation Functions for Eta Mesons,''
Phys.\ Rev.\ D {\bf 83} (2011) 034002
doi:10.1103/PhysRevD.83.034002
[arXiv:1009.6145 [hep-ph]].
%%CITATION = doi:10.1103/PhysRevD.83.034002;%%

%%%%%%%%%%%%%%%%%%%%%%%%%%%%% FFs_DGLAP %%%%%%%%%%%%%%%%%%%%%%%%%%%%%%%%%%%%%%
%\cite{Indumathi:2011vn}
\bibitem{Indumathi:2011vn}
D.~Indumathi and H.~Saveetha,
%``Study of Vector Meson Fragmentation Using a Broken SU(3) Model,''
Int.\ J.\ Mod.\ Phys.\ A {\bf 27}, 1250103 (2012)
%doi:10.1142/S0217751X12501035
[arXiv:1102.5594 [hep-ph]].
%%CITATION = doi:10.1142/S0217751X12501035;%%

%\cite{Saveetha:2013jda}
\bibitem{Saveetha:2013jda}
H.~Saveetha, D.~Indumathi and S.~Mitra,
%``Vector meson fragmentation using a model with broken SU(3) at the Next-to-Leading Order,''
Int.\ J.\ Mod.\ Phys.\ A {\bf 29}, no. 07, 1450049 (2014)
%doi:10.1142/S0217751X14500493
[arXiv:1309.2134 [hep-ph]].
%%CITATION = doi:10.1142/S0217751X14500493;%%

%\cite{Hirai:2011si}
\bibitem{Hirai:2011si}
M.~Hirai and S.~Kumano,
%``Numerical solution of $Q^2$ evolution equations for fragmentation functions,''
Comput.\ Phys.\ Commun.\  {\bf 183}, 1002 (2012)
%doi:10.1016/j.cpc.2011.12.022
[arXiv:1106.1553 [hep-ph]].
%%CITATION = doi:10.1016/j.cpc.2011.12.022;%%
%%%%%%%%%%%%%%%%%%%%%%%%%%%%% FFs_DGLAP %%%%%%%%%%%%%%%%%%%%%%%%%%%%%%%%%%%%%%

%%%%%%%%%%%%%   AKK08, KKP, KRE, AESSS  %%%%%%%%%%%%%%%%
%\cite{Kniehl:2008et}
\bibitem{Kniehl:2008et}
B.~A.~Kniehl,
%``Status of AKK Fragmentation Functions,''
doi:10.3360/dis.2008.152
arXiv:0807.2214 [hep-ph].
%%CITATION = doi:10.3360/dis.2008.152;%%

%\cite{Kretzer:2000yf}
\bibitem{Kretzer:2000yf}
S.~Kretzer,
%``Fragmentation functions from flavor inclusive and flavor tagged e+ e- annihilations,''
Phys.\ Rev.\ D {\bf 62} (2000) 054001
doi:10.1103/PhysRevD.62.054001
[hep-ph/0003177].
%%CITATION = doi:10.1103/PhysRevD.62.054001;%%

%%%%%%%%%%%%%%%%%%  FFs uncertainty %%%%%%%%%%%%%%%%%%%%%%%
%\cite{deFlorian:2007aj}
\bibitem{deFlorian:2007aj}
D.~de Florian, R.~Sassot and M.~Stratmann,
%``Global analysis of fragmentation functions for pions and kaons and their uncertainties,''
Phys. Rev. D \textbf{75} (2007), 114010
doi:10.1103/PhysRevD.75.114010
[arXiv:hep-ph/0703242 [hep-ph]].

%\cite{dEnterria:2013sgr}
\bibitem{dEnterria:2013sgr}
D.~d'Enterria, K.~J.~Eskola, I.~Helenius and H.~Paukkunen,
%``Confronting current NLO parton fragmentation functions with inclusive charged-particle spectra at hadron colliders,''
Nucl. Phys. B \textbf{883} (2014), 615-628
doi:10.1016/j.nuclphysb.2014.04.006
[arXiv:1311.1415 [hep-ph]].

%\cite{Metz:2016swz}
\bibitem{Metz:2016swz}
A.~Metz and A.~Vossen,
%``Parton Fragmentation Functions,''
Prog. Part. Nucl. Phys. \textbf{91} (2016), 136-202
doi:10.1016/j.ppnp.2016.08.003
[arXiv:1607.02521 [hep-ex]].

%%%%%%%%%%%%%%%%%   pp_h   %%%%%%%%%%%%%%%%%%%%%%%
%\cite{CMS:2012aa}
\bibitem{CMS:2012aa}
S.~Chatrchyan \textit{et al.} [CMS],
%``Study of high-pT charged particle suppression in PbPb compared to $pp$ collisions at $\sqrt{s_{NN}}=2.76$~TeV,''
Eur. Phys. J. C \textbf{72} (2012), 1945
doi:10.1140/epjc/s10052-012-1945-x
[arXiv:1202.2554 [nucl-ex]].

%\cite{CMS:2016xef}
\bibitem{CMS:2016xef}
V.~Khachatryan \textit{et al.} [CMS],
%``Charged-particle nuclear modification factors in PbPb and pPb collisions at $ \sqrt{s_{\mathrm{N}\;\mathrm{N}}}=5.02 $~TeV,''
JHEP \textbf{04} (2017), 039
doi:10.1007/JHEP04(2017)039
[arXiv:1611.01664 [nucl-ex]].
%%%%%%%%%%%%%%%%%   pp_h   %%%%%%%%%%%%%%%%%%%%%%%

%%%%%%%%%%%%%%%%%   pp_pi0,eta   %%%%%%%%%%%%%%%%%%%%%%%
%\cite{Acharya:2017hyu}
\bibitem{Acharya:2017hyu}
S.~Acharya \textit{et al.} [ALICE],
%``Production of ${\pi ^0}$ and $\eta $ mesons up to high transverse momentum in pp collisions at 2.76~TeV,''
Eur. Phys. J. C \textbf{77} (2017) no.5, 339
doi:10.1140/epjc/s10052-017-4890-x
[arXiv:1702.00917 [hep-ex]].

%\cite{ALICE:2019hno}
\bibitem{ALICE:2019hno}
S.~Acharya \textit{et al.} [ALICE],
%``Production of charged pions, kaons, and (anti-)protons in Pb-Pb and inelastic $pp$ collisions at $\sqrt {s_{NN}}$ = 5.02~TeV,''
Phys. Rev. C \textbf{101} (2020) no.4, 044907
doi:10.1103/PhysRevC.101.044907
[arXiv:1910.07678 [nucl-ex]].

%\cite{Abelev:2012cn}
\bibitem{Abelev:2012cn}
B.~Abelev \textit{et al.} [ALICE],
%``Neutral pion and $\eta$ meson production in proton-proton collisions at $\sqrt{s}=0.9$~TeV and $\sqrt{s}=7$~TeV,''
Phys. Lett. B \textbf{717} (2012), 162-172
doi:10.1016/j.physletb.2012.09.015
[arXiv:1205.5724 [hep-ex]].

%\cite{Acharya:2017tlv}
\bibitem{Acharya:2017tlv}
S.~Acharya \textit{et al.} [ALICE],
%``$\pi ^{0}$ and $\eta $ meson production in proton-proton collisions at $\sqrt{s}=8$~TeV,''
Eur. Phys. J. C \textbf{78} (2018) no.3, 263
doi:10.1140/epjc/s10052-018-5612-8
[arXiv:1708.08745 [hep-ex]].
%%%%%%%%%%%%%%%%%   pp_pi0,eta   %%%%%%%%%%%%%%%%%%%%%%%

%%%%%%%%%%%%%%%%%   pp_phi   %%%%%%%%%%%%%%%%%%%%%%%
%\cite{Adam:2017zbf}
\bibitem{Adam:2017zbf}
J.~Adam \textit{et al.} [ALICE],
%``K$^{*}(892)^{0}$ and $\phi(1020)$ meson production at high transverse momentum in pp and Pb-Pb collisions at $\sqrt{s_\mathrm{NN}}$ = 2.76~TeV,''
Phys. Rev. C \textbf{95} (2017) no.6, 064606
doi:10.1103/PhysRevC.95.064606
[arXiv:1702.00555 [nucl-ex]].

%\cite{ALICE:2021ptz}
\bibitem{ALICE:2021ptz}
S.~Acharya \textit{et al.} [ALICE],
%``Production of K$^{*}(892)^{0}$ and $\phi(1020)$ in pp and Pb-Pb collisions at $\sqrt{s_{\rm NN}} = 5.02$~TeV,''
[arXiv:2106.13113 [nucl-ex]].

%\cite{ALICE:2019hyb}
\bibitem{ALICE:2019hyb}
S.~Acharya \textit{et al.} [ALICE],
%``$\rm{K}^{*}(\rm{892})^{0}$ and $\phi(1020)$ production at midrapidity in pp collisions at $\sqrt{s}$ = 8~TeV,''
Phys. Rev. C \textbf{102} (2020) no.2, 024912
doi:10.1103/PhysRevC.102.024912
[arXiv:1910.14410 [nucl-ex]].

%\cite{ALICE:2020jsh}
\bibitem{ALICE:2020jsh}
S.~Acharya \textit{et al.} [ALICE],
%``Production of light-flavor hadrons in pp collisions at $\sqrt{s}~=~7\text { and }\sqrt{s} = 13 \, \text {~TeV} $,''
Eur. Phys. J. C \textbf{81} (2021) no.3, 256
doi:10.1140/epjc/s10052-020-08690-5
[arXiv:2005.11120 [nucl-ex]].
%%%%%%%%%%%%%%%%%   pp_phi 

%%%%%%%%%%%%%%     GM     %%%%%%%%%%%%%%%%%%%%%%%%%%
%\cite{Moller:2015fba}
\bibitem{Moller:2015fba}
P.~M\"oller, A.~J.~Sierk, T.~Ichikawa and H.~Sagawa,
%``Nuclear ground-state masses and deformations: FRDM(2012),''
Atom. Data Nucl. Data Tabl. \textbf{109-110} (2016), 1-204
doi:10.1016/j.adt.2015.10.002
[arXiv:1508.06294 [nucl-th]].

%\cite{Zakharov:2018ctz}
\bibitem{Zakharov:2018ctz}
B.~G.~Zakharov,
%``Monte Carlo Glauber model with meson cloud: predictions for 5.44 TeV Xe + Xe collisions,''
Eur. Phys. J. C \textbf{78} (2018) no.5, 427
doi:10.1140/epjc/s10052-018-5924-8
[arXiv:1804.05405 [nucl-th]].
%%%%%%%%%%%%%%     GM     %%%%%%%%%%%%%%%%%%%%%%%%%%

%%%%%%%%%%%%%%%%%%%%   CLV
%\cite{Pang:2012he}
\bibitem{Pang:2012he}
L.~Pang, Q.~Wang and X.~Wang,
%``Effects of initial flow velocity fluctuation in event-by-event (3+1)D hydrodynamics,''
Phys. Rev. C \textbf{86} (2012), 024911
doi:10.1103/PhysRevC.86.024911
[arXiv:1205.5019 [nucl-th]].

%\cite{Pang:2014ipa}
\bibitem{Pang:2014ipa}
L.~G.~Pang, Y.~Hatta, X.~N.~Wang and B.~W.~Xiao,
%``Analytical and numerical Gubser solutions of the second-order hydrodynamics,''
Phys. Rev. D \textbf{91} (2015) no.7, 074027
doi:10.1103/PhysRevD.91.074027
[arXiv:1411.7767 [hep-ph]].

%\cite{Pang:2018zzo}
\bibitem{Pang:2018zzo}
L.~G.~Pang, H.~Petersen and X.~N.~Wang,
%``Pseudorapidity distribution and decorrelation of anisotropic flow within the open-computing-language implementation CLVisc hydrodynamics,''
Phys.\ Rev.\ C {\bf 97} (2018) no.6,  064918
doi:10.1103/PhysRevC.97.064918
[arXiv:1802.04449 [nucl-th]].
%%%%%%%%%%%%%%%%%%%%   CLV
%%%%%%%%%%%%%%     hydro     %%%%%%%%%%%%%%%%%%%%%%%%%%

%\cite{Eskola:2016oht}
\bibitem{Eskola:2016oht}
K.~J.~Eskola, P.~Paakkinen, H.~Paukkunen and C.~A.~Salgado,
%``EPPS16: Nuclear parton distributions with LHC data,''
Eur.\ Phys.\ J.\ C \textbf{77} (2017) no.3, 163
doi:10.1140/epjc/s10052-017-4725-9
[arXiv:1612.05741 [hep-ph]].

%%%%%%%%%%%%%%%%%%%    ALICE h RAA  %%%%%%%%%%%%%%%%%%%%%%%%%%
%\cite{ALICE:2018vuu}
\bibitem{ALICE:2018vuu}
S.~Acharya \textit{et al.} [ALICE],
%``Transverse momentum spectra and nuclear modification factors of charged particles in pp, p-Pb and Pb-Pb collisions at the LHC,''
JHEP \textbf{11} (2018), 013
doi:10.1007/JHEP11(2018)013
[arXiv:1802.09145 [nucl-ex]].
%%%%%%%%%%%%%%%%%%%%%   ALICE h RAA

%\cite{Moreau:2019vhw}
\bibitem{Moreau:2019vhw}
P.~Moreau, O.~Soloveva, L.~Oliva, T.~Song, W.~Cassing and E.~Bratkovskaya,
%``Exploring the partonic phase at finite chemical potential within an extended off-shell transport approach,''
Phys. Rev. C \textbf{100} (2019) no.1, 014911
doi:10.1103/PhysRevC.100.014911
[arXiv:1903.10257 [nucl-th]].

%\cite{Grishmanovskii:2022tpb}
\bibitem{Grishmanovskii:2022tpb}
I.~Grishmanovskii, T.~Song, O.~Soloveva, C.~Greiner and E.~Bratkovskaya,
%``Exploring jet transport coefficients in the strongly interacting quark-gluon plasma,''
[arXiv:2204.01561 [nucl-th]].
%%%%%%%%%%%%%%%%%%%%%%  qhat


\end{thebibliography}
\end{document}